\documentclass[12pt]{iopart}
\usepackage[dvips]{graphicx}
\usepackage{amssymb}
\usepackage{amsfonts}
\newcommand {\dr}{{\mathrm d}\mathbf{r}}
\newcommand {\dR}{{\mathrm d}\mathbf{R}}
\newcommand {\rr}{\mathbf{r}}
\newcommand {\RR}{\mathbf{R}}
\newcommand {\dd}{{\mathrm d}}
\newcommand {\dpp}{{\mathrm d}\mathbf{p}}
\newcommand {\dP}{{\mathrm d}\mathbf{P}}
\newcommand {\pp}{\mathbf{p}}
\newcommand {\PP}{\mathbf{P}}
\newcommand {\XX}{\mathbf{X}}
\newcommand {\xx}{\mathbf{x}}
\newcommand {\YY}{\mathbf{Y}}
\newcommand {\ZZ}{\mathbf{Z}}
\newcommand {\JJ}{\mathbf{J}}
\newcommand {\GG}{\mathbf{G}}
\newcommand {\jj}{\mathbf{j}}

\begin{document}

\title[DDFT: binary phase separating colloidal fluid in a cavity]
{Dynamical density functional theory:\\
binary phase-separating colloidal fluid in a cavity}

\author{A J Archer\footnote{andrew.archer@bristol.ac.uk}} 
\address{H H Wills Physics Laboratory, University of Bristol,
Bristol BS8 1TL, UK}
\date{\today}

\begin{abstract}
The dynamical density functional theory of Marconi and
Tarazona [{\it J. Chem. Phys.}, {\bf 110}, 8032 (1999)], a theory for
the non-equilibrium dynamics of the
one-body density profile of a colloidal fluid, is applied
to a binary fluid mixture of repulsive Gaussian particles
confined in a spherical cavity of variable size.
For this model fluid there exists
an extremely simple Helmholtz free energy functional
that provides a remarkably accurate description of the equilibrium fluid
properties.
We therefore use this functional to test the assumptions implicit in the
dynamical density functional theory, rather than any approximations involved in
constructing the free energy functional. We find very good agreement between the
theory and Brownian dynamics simulations, focusing on cases
where the confined fluid exhibits phase separation in the cavity.
We also present an instructive derivation of the Smoluchowski equation (from
which one is able to derive the dynamical density functional theory) starting
from the Liouville equation -- a fully microscopic treatment of the colloid
{\em and} solvent particles. This `coarse-graining' is, of course, not exact and
thus the derivation demonstrates the physical assumptions implicit in the
Smoluchowski equation and therefore also in the dynamical density functional
theory.
\end{abstract}


\section{Introduction}
For a multi-component
fluid composed of particles with a large size difference between the
different components, a theoretical description of the (inhomogeneous)
fluid dynamics is challenging due to the varying equilibration timescales for
the different species of particles in the fluid.
Colloidal suspensions are a particular example of such complex fluids.
Typically, the colloids are 10--1000nm in diameter, whereas the solvent
molecules are, in the case of water, $\sim 0.2$nm.
The situations in which a theory could be used to describe the motion of
inhomogeneous (confined) colloidal particles are many. For example, in
colloidal physics the use of optical tweezers to confine and then move
individual of groups of colloids is common \cite{opticaltweezers}.
Simulating such a fluid is also computationally expensive, because
of the huge numbers of solvent molecules required when simulating even a limited
number of colloids. A common approach to such systems is to integrate out the
degrees of freedom of the smaller particles in order to render a description of
the system based only on the (much smaller) phase space corresponding to the
degrees of freedom of the big colloid particles \cite{BH}.

In the case of colloidal fluids in thermal (static) equilibrium,
this `coarse-grained' approach is, at least
formally, well understood and one is able to formulate an effective
Hamiltonian for the fluid involving only the phase space coordinates of
the big particles. Of course, the effective potential between the
big particles is, in general, many-body in character and dependent on the
density of both the solvent and colloid particles
\cite{BH,Likos}. For example, the depletion potential, or effective solvent
mediated interaction potential, between big hard-sphere particles dissolved in a
solvent of small hard spheres, is oscillatory for large separations
\cite{Roland} and is therefore very different from the `bare' potentials one
typically encounters between the particles of
a molecular fluid. However, for the non-equilibrium
dynamics of colloidal fluids, coarse-graining is carried
out on a more ad-hoc basis. A
common approach is to use the effective solvent mediated potential obtained
from the equilibrium theory together with stochastic
equations of motion -- i.e.\ the force on the colloids due to collisions with
the solvent particles is modelled as a random white noise term and a frictional
(Stokes) one-body drag force. The
colloids are thus modelled as Brownian particles.
From the (Langevin) equations of motion for the Brownian particles
one can obtain the Fokker-Planck (Smoluchowski) equation \cite{risken84} for
the colloid probability distribution function in phase space, and thus one is
able to determine non-equilibrium
dynamic properties of the colloidal fluid. In going to such a
stochastic description of the dynamics one inevitably neglects hydrodynamic
effects. We justify the above stochastic approach in Sec.\ \ref{sec:LKS}. 

In this paper we consider cases of a model colloidal fluid confined in
time dependent one-body external potentials. We focus on the time dependence of
the fluid one-body density profile, using the dynamical density functional
theory (DDFT) derived recently by Marconi and Tarazona
\cite{Marconi:TarazonaJCP1999,Marconi:TarazonaJPCM2000} \footnote{The key DDFT
equation (Eq.\ (\ref{eq:mainres}))
was proposed originally, without derivation, by Evans \cite{Evans79}.}.
In a recent paper \cite{Archer7} (see also Ref.\ \cite{Archer8}),
it was demonstrated that this DDFT
could be derived from the Smoluchowski equation, by making the assumption that
the form of
correlations between particles in an inhomogeneous fluid out of equilibrium,
is the same as in an equilibrium fluid with the same one body density
profile \cite{Marconi:TarazonaJCP1999,Marconi:TarazonaJPCM2000,Archer7}.
We apply the DDFT to the dynamics of a particular model colloidal fluid, the
binary Gaussian core model (GCM), making a comparison between the results from
the DDFT and Brownian dynamics (BD) simulation results. Dzubiella and Likos
\cite{joe:christos} made a similar comparison for the one component GCM, and
found good agreement between theory and simulations. The GCM is a simple model
for the effective potential between the centers of mass of polymers or
dendrimers in solution \cite{Likos, Dautenhahn, LouisetalPRL2000,
BolhuisetalJCP2001, LouisetalPhysicaA2002, BolhuisLouisMacrom2002, cnl:jcp,
cnl:macrom}. The reason Dzubiella and Likos chose this
particular model fluid is that there exists a very simple, yet remarkably
accurate (as we demonstrate below for mixtures) Helmholtz free energy functional
for describing the equilibrium properties of the GCM
\cite{Likos,LangJPCM,paper1,Archer1,Archer2,Archer6,Patrykiejewetal2004} and
thus by using this functional combined with the DDFT they were able to test some
of the
approximations inherent in the DDFT, rather than the approximations involved in
the free energy functional.
The DDFT has also been successfully applied to situations with steady currents
\cite{flor2,flor1} and very recently, Rex
\etal~have applied the DDFT to the GCM fluid in shear and travelling wave
potentials \cite{RexetalPRE2005}. For these cases they also find good agreement
between theory and simulations.
One particular question we wish to address here is whether the DDFT can
incorporate dynamical effects that are specific to colloidal fluids
composed of two different species of colloids and, in particular, dynamical
effects arising when there is phase separation and wetting phenomena in a
confined colloidal fluid.

The paper is arranged as follows: In Sec.\ \ref{sec:theDDFT} we introduce and
briefly describe the DDFT. The results from the application of the DDFT are
presented in Sec.\ \ref{sec:application}. 
In Sec.\ \ref{sec:LKS} we give an alternative derivation of the
Smoluchowski equation. Starting from the Liouville equations for a
mixture of colloid and solvent particles, we first obtain the Kramers equation
and then from this the Smoluchowski equation. These equations can also be
derived as the (generalised) Fokker-Planck equations for a
system of Brownian particles with stochastic equations of motion. The present
derivation elucidates some of the physical assumptions concerning the fluid
dynamics that are implicit in the Kramers and the Smoluchowski equations. We
believe this derivation sheds light on the status of the DDFT and the
approximations inherent in this theory. Finally, in Sec.\ \ref{sec:discussion}
we discuss our results and draw some conclusions.

\section{The DDFT}
\label{sec:theDDFT}

Before introducing the DDFT, we recall briefly some of the basic results from
{\em equilibrium} density functional theory (EDFT) \cite{Evans79,Evans92} that
are required for our discussion of the DDFT.
The key quantity in EDFT is the Helmholtz free energy functional:
\begin{eqnarray}
F[\rho(\rr)] = F_{id}[\rho(\rr)]
+F_{ex}[\rho(\rr)]+\int \dr u_{ext}(\rr) \rho(\rr),
\label{eq:freeenergy}
\end{eqnarray}
which is a functional of $\rho(\rr)$, the fluid one-body density. For simplicity
we consider a one-component fluid. $u_{ext}(\rr)$
is a one body external potential and
\begin{eqnarray}
F_{id}[\rho(\rr)] =k_BT \int \dr \rho(\rr)[\ln(\Lambda^3\rho(\rr))-1],
\label{eq:idgas_freeenergy}
\end{eqnarray}
is the (exact) ideal gas free energy; $\Lambda$ is the thermal de Broglie
wavelength of the particles.
$F_{ex}[\rho(\rr)]$ is the excess (over ideal) contribution to the free
energy due to interactions between the particles \cite{Evans79,Evans92}. In EDFT
one considers the following grand potential functional
\begin{eqnarray}
\Omega[\rho(\rr)] = F[\rho(\rr)]- \mu \int \dr \rho(\rr),
\label{eq:grand_pot}
\end{eqnarray}
where $\mu$ is the chemical potential. The equilibrium density
distribution is given by the minimisation condition \cite{Evans79,Evans92}:
\begin{eqnarray}
\frac{\delta \Omega}{ \delta \rho(\rr)}=0.
\label{eq:min_cond}
\end{eqnarray}
Thus, using Eqs.\ (\ref{eq:freeenergy}) -- (\ref{eq:min_cond}),
one obtains the following Euler-Lagrange equation for the equilibrium fluid
density profile:
\begin{eqnarray}
0=k_BT \ln \Lambda^3 \rho(\rr) - k_BT c^{(1)}(\rr)+u_{ext}(\rr)-\mu,
\label{eq:EL_eq}
\end{eqnarray}
where the one body direct correlation function:
\begin{equation}
c^{(1)}(\rr) \equiv -\beta \frac{\delta F_{ex}[\rho(\rr)]}{\delta \rho(\rr)}.
\label{eq:c1}
\end{equation}
$\beta=1/k_BT$ is the inverse temperature and
$-k_BTc^{(1)}(\rr)$ is an effective one body potential that incorporates
the effect of the interparticle interactions in the fluid.
The central task in EDFT is to find a suitable approximation for
$F_{ex}[\rho(\rr)]$ for the fluid of interest. $F_{ex}[\rho(\rr)]$
is, in general, an unknown quantity. 
A second functional derivative of $F_{ex}[\rho(\rr)]$ gives the inhomogeneous
(Ornstein-Zernike) pair direct correlation function:
\begin{equation}
c^{(2)}(\rr,\rr') \equiv
-\beta \frac{\delta^2 F_{ex}[\rho(\rr)]}
{\delta \rho(\rr) \delta \rho(\rr')}.
\label{eq:c2}
\end{equation}
Often, approximations for $F_{ex}[\rho(\rr)]$ are constructed by requiring that
the approximate excess Helmholtz free energy functional generate, via Eq.\
(\ref{eq:c2}), reliable results for $c^{(2)}(r)$, the {\em bulk} pair direct
correlation function, about which much is known from the theory of bulk
equilibrium fluids \cite{Evans92,HM}.

We now move on to
consider a non-equilibrium fluid of $N$ Brownian colloid particles.
We denote the position of the $i^{{\mathrm th}}$ colloid particle
by $\rr_i$, and the set of all position coordinates by
$\rr^N \equiv \{\rr_1, ...,\rr_N\}$. The total colloid
potential energy $U_N(\rr^N,t)$ is assumed to be of the following form:
\begin{eqnarray}
\fl
U_N(\rr^N,t)
=\sum_{i=1}^N u_{ext}(\rr_i,t) 
+ \frac{1}{2} \sum_{j \neq i} \sum_{i=1}^N u_2(\rr_i,\rr_j) 
+ \frac{1}{6} \sum_{k \neq j \neq i}\sum_{j \neq i} \sum_{i=1}^N
u_3(\rr_i,\rr_j,\rr_k)  + ...,
\label{eq:U_fn}
\end{eqnarray}
which is made up of a one-body term, i.e.\ a {\em time-dependent} one-body
external potential
$u_{ext}(\rr_i,t)$ acts on each fluid particle; a two-body
term, which is a sum of (time independent)
pair potentials $u_2(\rr_i,\rr_j)$; a three body term,
given by a sum of three-body potentials $u_3(\rr_i,\rr_j,\rr_k)$ and higher body
terms, each given by a sum of multi-body potentials.
Since we do not include explicitly the solvent particles,
this potential must include contributions from solvent mediated interactions
between the colloids as well as the direct (bare) interactions between the
colloids -- we will return to this issue in Sec.\ \ref{sec:LKS}\footnote{For the
purposes of calculating equilibrium fluid properties,
all solvent effects are included in the effective potential $U_N(\rr^N,t)$.
However, for {\em dynamics}, the solvent friction has to be explicitly taken
into account.}. On the Brownian time scale the equations of motion of
the colloids obey the following stochastic (Langevin) equations
of motion \cite{Dhont_book}:
\begin{equation}
\Gamma^{-1}\frac{\dd \rr_i(t)}{\dd t}
=-\frac{ \partial U_N(\rr^N,t)}{\partial \rr_i}+\GG_i(t),
\label{eq:langevin2}
\end{equation}
where $\Gamma^{-1}$ is a friction constant characterising the one-body drag of
the solvent on the colloidal particles and
$\GG_i(t)=(\xi_i^x(t),\xi_i^y(t),\xi_i^z(t))$ is a white noise term
with the property
\begin{eqnarray}
\left< \xi_i^{\alpha}(t) \right> =0, \nonumber \\
\left< \xi_i^{\alpha}(t)\xi_i^{\nu}(t')\right> = 2 k_BT \delta_{ij}
\delta^{\alpha \nu} \delta(t-t').
\label{eq:noise_term}
\end{eqnarray}
The stochastic equations of motion (\ref{eq:langevin2}), of course, neglect
hydrodynamic interactions between the colloids.

We can define a probability density function $P^{(N)}(\rr^N,t)$ for the $N$
colloids. The time evolution of $P^{(N)}(\rr^N,t)$ is described by the
Smoluchowski equation
\cite{risken84}:
\begin{eqnarray}
\frac{\partial P^{(N)}}{\partial t} = \Gamma \sum_{i=1}^N
\frac{\partial}{\partial \rr_i} \cdot
\left( k_BT \frac{\partial P^{(N)}}{\partial \rr_i}+
\frac{\partial U_N}{\partial \rr_i}P^{(N)} \right).
\label{eq:Smol}
\end{eqnarray}
This equation is the (generalised) Fokker-Planck equation for the
Langevin equations (\ref{eq:langevin2}).
The Smoluchowski equation is generally presented from this stochastic viewpoint.
However, as we show below in Sec.\ \ref{sec:LKS}, one can argue for its use
as an approximation to the exact Liouville equations:
By going to the Smoluchowski equation the
description of the fluid is reduced to one based solely on the position
coordinates of the colloids, rather than utilising the full set of phase space
coordinates for the colloid and solvent particles. However, in practice
a further reduction
is required in order to be able to determine explicitly the dynamics of the
colloids. In particular, we focus on the colloid one-body density
\cite{Archer7,Archer8,HM}:
\begin{equation}
\rho(\rr_1,t)  = N \int \dr_2 \, ... \int \dr_N
P^{(N)}(\rr^N,t).
\label{eq:rho_1}
\end{equation}
Similarly, the $n$-particle density is
\begin{equation}
\rho^{(n)}(\rr^n,t)  = \frac{N!}{(N-n)!} \int \dr_{n+1} \,
...\int \dr_N P^{(N)}(\rr^N,t).
\label{eq:rho_n}
\end{equation}
We now follow closely the derivation in Ref.\ \cite{Archer7}:
Using Eqs.\ (\ref{eq:U_fn}), (\ref{eq:rho_1}) and (\ref{eq:rho_n}),
we find that on integrating Eq.\ (\ref{eq:Smol}), one obtains
\cite{Archer7,DhontJCP1996}:
\begin{eqnarray}
\fl
\nonumber
\Gamma^{-1} \frac{\partial \rho(\rr_1,t)}{\partial t}  = k_BT
\frac{\partial^2 \rho(\rr_1,t)}{\partial \rr_1^2}
+ \frac{\partial}{\partial \rr_1} \cdot \left[\rho(\rr_1,t) \frac{\partial
u_{ext}(\rr_1,t)}{\partial \rr_1} \right] \nonumber \\
+ \frac{\partial}{\partial \rr_1} \cdot \int \dr_2 \rho^{(2)}(\rr_1,\rr_2,t)
\frac{\partial u_2(\rr_1,\rr_2)}{\partial \rr_1} \nonumber \\
+\frac{\partial}{\partial \rr_1} \cdot\int \dr_2 \int \dr_3
\rho^{(3)}(\rr_1,\rr_2,\rr_3,t)
\frac{\partial u_3(\rr_1,\rr_2,\rr_3)}{\partial \rr_1}
+  ....
\label{eq:Dhont_eq}
\end{eqnarray}
At equilibrium, when $\partial \rho(\rr,t)/ \partial t=0$, this equation is just
the gradient of the (exact) first equation of the YBG hierarchy \cite{HM}.

We are now in a position to make contact with EDFT.
For an {\em equilibrium} fluid the gradient of $-  k_BT c^{(1)}(\rr)$
(see Eq.\ (\ref{eq:c1})), a one body force due to interactions between particles
in the fluid, is given by the following sum-rule
\cite{Evans79,Archer7,RowlinsonWidom}:
\begin{eqnarray}
-  k_BT \rho(\rr_1) \frac{\partial c^{(1)}(\rr_1)}{\partial \rr_1}
= \sum_{n=2}^{\infty} \int \dr_2 ... \int \dr_n \rho^{(n)}(\rr^n)
\frac{\partial u_n(\rr^n)}{\partial \rr_1}.
\label{eq:grad_c1_many}
\end{eqnarray}
If we assume that we can use (\ref{eq:grad_c1_many}) for a non-equilibrium fluid
we are assuming that the effective one body force on a particle in the fluid due
to interactions with the other particles is the same as that in the equilibrium
fluid with the {\em same} one body density profile. We make this assumption and
using Eqs.\ (\ref{eq:grad_c1_many}) and (\ref{eq:freeenergy}) together with Eq.\
(\ref{eq:Dhont_eq}) we obtain the key DDFT equation
\cite{Marconi:TarazonaJCP1999,Marconi:TarazonaJPCM2000,Archer7}:
\begin{equation}
\frac{\partial \rho(\rr,t)}{\partial t}
=\Gamma \frac{\partial}{\partial \rr} \cdot
\left[ \rho(\rr,t) \frac{\partial}{\partial \rr}
\left(\frac{\delta F[\rho(\rr,t)]}{\delta
\rho(\rr,t)}\right)  \right],
\label{eq:mainres}
\end{equation}
where $F[\rho(\rr,t)]$ is given by Eq.\ (\ref{eq:freeenergy}) with $\rho(\rr)$
replaced by the time dependent one-body density $\rho(\rr,t)$. Before commenting
on the status of
Eq.\ (\ref{eq:mainres}), we note that the above arguments can easily be
generalised to the case where there are several different species of colloids.
If there are $Q$ different species of colloids, with $N_q$ colloids of
species $q$, such that the total number of colloidal particles is
$N=\sum_{q=1}^Q N_q$, and the total colloid effective potential energy is
(c.f.\ Eq.\ (\ref{eq:U_fn})):
\begin{eqnarray}
U_N(\rr^N,t)  = \sum_{q=1}^Q\sum_{i=1}^{N_Q} u_{ext}^q(\rr_{q,i},t) 
+\frac{1}{2} \sum_{q,q'=1}^Q\sum_{j \neq i} \sum_{i=1}^{N_Q}
u_2^{q,q'}(\rr_{q,i},\rr_{q',j})
+\, ...
\label{eq:U_fn_Q}
\end{eqnarray}
In this case, the Smoluchowski equation (\ref{eq:Smol}) simply becomes
\begin{eqnarray}
\frac{\partial P^{(N)}}{\partial t} = \sum_{q=1}^Q \Gamma_q \sum_{i=1}^{N_q}
\frac{\partial}{\partial \rr_{q,i}} \cdot
\left( k_BT \frac{\partial P^{(N)}}{\partial \rr_{q,i}}+
\frac{\partial U_N}{\partial \rr_{q,i}}P^{(N)} \right),
\label{eq:Smol_Ncomp}
\end{eqnarray}
where $\Gamma_q^{-1}$ is the friction constant for the $q^{\mathrm th}$ species
of colloid particle. One can then integrate Eq.\ (\ref{eq:Smol_Ncomp}) to obtain
a DDFT for the one-body density profiles. The one-body density of species $q$
is:
\begin{equation}
\rho_q(\rr_{q,1},t)  = N_q \prod_{i=2}^{N_q}\int \dr_{q,i}  \prod_{q'
\neq q} \int \dr^{N_{q'}} P^{(N)}(\rr^N,t),
\label{eq:rho_q1}
\end{equation}
i.e., $\rho_q(\rr_{q,1},t)$ is given by the integral over $P^{(N)}$ with respect
to all the colloid position coordinates other than those of the $i=1$ colloid of
species $q$. The multi-component generalisation of the DDFT equation
(\ref{eq:mainres}) then becomes:
\begin{equation}
\frac{\partial \rho_q(\rr,t)}{\partial t}
=\Gamma_q \frac{\partial}{\partial \rr} \cdot
\left[ \rho_q(\rr,t) \frac{\partial}{\partial \rr}
\left(\frac{\delta F[\{ \rho_q(\rr,t) \}]}{\delta
\rho_q(\rr,t)}\right)  \right],
\label{eq:mainres_multi}
\end{equation}
where the Helmholtz free energy functional
for the multi-component colloidal fluid is (c.f.\ Eq.\ (\ref{eq:freeenergy})):
\begin{eqnarray}
F[\{ \rho_q(\rr,t) \}]&=&
\sum_{q=1}^Q k_BT \int \dr \rho_q(\rr,t)[\ln(\Lambda_q^3\rho_q(\rr,t))-1]
+F_{ex}[\{ \rho_q(\rr,t) \}] \nonumber \\
&+&\sum_{q=1}^Q \int \dr u_{ext}^q(\rr,t) \rho_q(\rr,t).
\label{eq:freeenergy_multi}
\end{eqnarray}

For an inhomogeneous
equilibrium fluid Eqs.\ (\ref{eq:grand_pot}) and (\ref{eq:min_cond})
imply that the chemical potential $\mu=\delta F/\delta \rho(\rr)$ is a constant
throughout the body of the fluid. Similarly, for an inhomogeneous equilibrium
multi-component fluid, the chemical potentials for the different species,
\begin{eqnarray}
\mu_q=\frac{\delta F[\{ \rho_q(\rr) \}]}{\delta \rho_q(\rr)},
\label{eq:mu_q}
\end{eqnarray}
take a constant value throughout the fluid. The DDFT (\ref{eq:mainres_multi}) is
equivalent to assuming that in the non-equilibrium fluid, this is not the case,
and that the gradients of the chemical potentials are the thermodynamic forces
driving particle currents $\jj_q$ of each species \cite{Archer7,Evans79}:
\begin{eqnarray}
\jj_q(\rr,t)=-\Gamma_q \rho_q(\rr,t) \frac{\partial \mu_q}{\partial \rr}.
\label{eq:current_q}
\end{eqnarray}
On combining this result with the continuity equation
\begin{eqnarray}
\frac{\partial \rho_q(\rr,t)}{\partial t}
=-\frac{\partial}{\partial \rr} \cdot \jj_q(\rr,t),
\label{eq:continuity_2}
\end{eqnarray}
one obtains the DDFT Eq.\ (\ref{eq:mainres_multi}). The DDFT is clearly not a
theory for a colloidal fluid in which there is a temperature gradient. In order
to incorporate such an effect in a microscopic theory, one would need to
construct an external potential that couples to both the position {\em and}
momentum degrees of freedom of the colloids. In reality, such thermal effects
normally also influence
the solvent particles. In the present description we have effectively
assumed that the solvent acts as a heat bath, keeping the temperature of the
colloids at a constant value -- even in cases where the change in the external
potential is such that one would find an increase in the temperature of the
fluid were it a simple (molecular, non-colloidal) fluid -- for example under
rapid compression.

As we have already mentioned, $F_{ex}[\{ \rho_q(\rr) \}]$, the
equilibrium excess Helmholtz free energy functional for the fluid mixture
is, in principle, an unknown quantity. There exist, depending
on the fluid in question, a number of accurate approximate
functionals, which give extremely good results for equilibrium fluid density
profiles for a wide variety of external potentials
(see, for example, Ref.\ \cite{Evans92} and references therein). This
means that the DDFT (\ref{eq:mainres_multi}) is a very appealing theory for the
dynamics of an {\em inhomogeneous} colloidal fluid,
since it builds directly upon EDFT, one of the most
successful theories for the equilibrium properties of inhomogeneous fluids. As
the presentation above shows, the DDFT is clearly not an exact theory.
However, the fact
that, in principle, the equilibrium profiles obtained from the DDFT are almost
exact leads one to expect that the theory should be reliable -- at least
when the fluid is not too far from equilibrium.

\section{Application of the DDFT to the GCM}
\label{sec:application}

We argued in the previous section that the DDFT (\ref{eq:mainres_multi}) should
be a good theory for the dynamics of a colloidal fluid when
the fluid is near to equilibrium, provided we have an accurate approximation for
the excess Helmholtz free energy functional $F_{ex}[\{ \rho_i(\rr,t) \}]$ for
the fluid (we now use $i,j$, rather than $q,q'$ to label the
different species of colloids). In this section we shall
demonstrate that in the case of a particular model fluid for which we {\em do}
have an accurate Helmholtz free energy
functional, the DDFT seems to be reliable even for situations when the
fluid is quite far from equilibrium and phenomena such as phase
separation and interfacial adsorption (wetting) are present. Throughout we
assume that
the Brownian level of description (\ref{eq:langevin2}) provides an accurate
account of the underlying particle dynamics.

Dzubiella and Likos \cite{joe:christos} applied the DDFT to a one-component
model fluid in
which the particles interact via a purely repulsive Gaussian potential, the
Gaussian core model (GCM). Their choice of model fluid was motivated by the fact
that a simple mean-field approximation for $F_{ex}[\{ \rho_i(\rr) \}]$ proves
to be quite accurate for the equilibrium properties,
and thus they were able to test whether the DDFT
formulation (\ref{eq:mainres}) of dynamics
is accurate. We shall say more about this model fluid
below. Their strategy was to consider the inhomogeneous fluid confined in either
a spherical cavity or a slit. They used EDFT to calculate the
equilibrium density profile corresponding to
a particular external potential and considered cases when
the external potential suddenly changed (i.e., a parameter in the external
potential was either increased or decreased).
Using the density profile from the EDFT as the starting density profile,
they used the DDFT to determine how the density
profile of the fluid evolved towards equilibrium. The reliability of
their results was assessed by making comparison with BD simulation
results -- i.e.\ they numerically integrated Eq.\ (\ref{eq:langevin2}) a number
of different times, for different realisations of the stochastic noise, and then
averaged over all the different runs
in order to obtain the ensemble average time evolution of the fluid
density profile. Dzubiella and Likos found that the DDFT and the BD
simulation results were in very good agreement \cite{joe:christos}. We
follow the same strategy here but for a binary mixture of GCM particles.

The one component GCM does not exhibit fluid-fluid phase separation
\cite{Likos,LangJPCM}. It is therefore of interest to find out whether the DDFT
proves to be reliable for the dynamics of inhomogeneous fluids when there
is phase separation in the fluid and when related interfacial
phenomena such as wetting of the cavity wall are present.
The model fluid we consider is the binary GCM. This fluid does exhibit
liquid-liquid phase separation \cite{paper1,Archer1}. The GCM
particles interact via the following purely repulsive pair potential:
\begin{equation}
u_{i,j}(r) = \epsilon_{i,j} \exp(-r^2/R_{i,j}^2)
\label{eq:pairpot}
\end{equation}
and no other higher body potentials. In Eq.\ (\ref{eq:pairpot}) $i,j=1,2$
label the two different species of particles, $\epsilon_{i,j}>0$ is a parameter
that determines the strength of the interaction and $R_{i,j}$ denotes
the range of the
interaction potential. Note that this potential has no hard core; the centres
of the particles can overlap completely. When one considers the effective
potential between the centres of mass of polymers in a good solvent, one finds
that the Gaussian potential (\ref{eq:pairpot}), with $\epsilon_{i,j} \sim
2k_BT$ and $R_{i,j} \sim R_g$, the polymer radius of gyration, provides a good
approximation \cite{Likos, Dautenhahn, LouisetalPRL2000, BolhuisetalJCP2001,
LouisetalPhysicaA2002, BolhuisLouisMacrom2002}. When the fluid density becomes
sufficiently high, each GCM particle interacts with a large number of
neighbours, and it is established that the following mean-field
excess Helmholtz free energy functional
\cite{Likos,LangJPCM,paper1,Archer1}:
\begin{equation}
F_{ex}[\{ \rho_i(\rr) \}]=\frac{1}{2}\sum_{i,j=1}^2 \int \dr \int \dr'
\rho_i(\rr) \rho_j(\rr')u_{i,j}(|\rr-\rr'|),
\label{eq:F_ex}
\end{equation}
becomes rather accurate. This functional generates the RPA closure
for the direct pair correlation functions (c.f.\ Eq.\ (\ref{eq:c2})):
\begin{equation}
c^{(2)}_{i,j}(\rr,\rr') \equiv
-\beta \frac{\delta^2 F_{ex}[\{ \rho_i(\rr) \}]}
{\delta \rho_i(\rr) \delta \rho_j(\rr')}
=-\beta u_{i,j}(|\rr-\rr'|).
\label{eq:c2_binary}
\end{equation}

We solve the DDFT (\ref{eq:mainres_multi}) for the binary GCM confined in
spherically symmetric external potentials of the form:
\begin{equation}
u_{ext}^i(r) = E \left(r/{\cal R} \right)^{10},
\label{eq:1}
\end{equation}
where $r$ is the distance from the origin, $E=10 k_BT$ and the length-scale
${\cal R}$ is the same for both species of particles.
This potential is of the same form as one of the external potentials considered
in Ref.\ \cite{joe:christos}
for a one component fluid of GCM particles. As the length parameter ${\cal R}$
is increased, the size of the cavity is increased. We consider cases
where for times $t<0$ the fluid is at equilibrium confined in a cavity with
potentials (\ref{eq:1}), with
${\cal R}={\cal R}_1$. Then at $t=0$ the cavity potentials change to ones with
${\cal R}={\cal R}_2 \neq {\cal R}_1$. Due to the spherical symmetry of the
external potentials, the (ensemble average) fluid one body density profiles will
also display spherical symmetry. In the binary GCM fluid we shall
consider mixtures with two different sets of pair potential parameters. The
first corresponds to a set giving bulk liquid-liquid phase separation and the
second to a set giving microphase-separation \cite{Archer6}. We shall also
assume throughout that the friction constants for the two
different species of particles are equal, i.e.,
$\Gamma^{-1}_1=\Gamma^{-1}_2=\Gamma^{-1}$, to keep the problem as simple as
possible.\footnote{Another more realistic choice, following Stokes, would be to
set $\Gamma_i^{-1} \propto R_{ii}$. However, we do not believe making this
alternative choice would affect any of our overall conclusions.}

\subsection{DDFT for a GCM fluid which exhibits bulk phase separation} 

We consider a binary GCM fluid with pair potential parameters
$\epsilon_{11}=\epsilon_{22}=2k_BT$,
$\epsilon_{12}=1.8877k_BT$, $R_{22}=0.665R_{11}$ and
$R_{12}=0.849R_{11}$, which exhibits bulk
fluid-fluid phase separation. The fact that
$R_{12}=(1+\Delta)(R_{11}+R_{22})/2$, with $\Delta>0$, ensures that the fluid
exhibits positive non-additivity and it is this feature
which induces phase separation. This
choice of pair potential parameters was
used in a number of previous studies by the
author \cite{Archer1,Archer2,Archer3,Archer5,Archer10}.
The fluid phase-separates at high
total densities $\rho=\rho_1+\rho_2$, where $\rho_1$ and $\rho_2$ are the bulk
densities of the two species. Within the mean-field DFT defined by Eq.\
(\ref{eq:F_ex}) the (lower) critical point is at $\rho
R_{11}^3=5.6$ and concentration $x \equiv \rho_2/ \rho =0.70$ \cite{Archer1}.

\begin{figure}
\begin{center}
\begin{minipage}[t]{5.1cm}
\includegraphics[width=5cm]{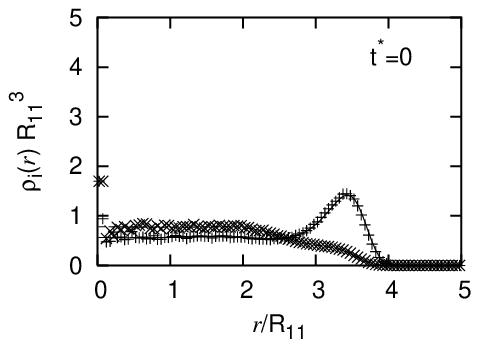}
\end{minipage}
\begin{minipage}[t]{5.1cm}
\includegraphics[width=5cm]{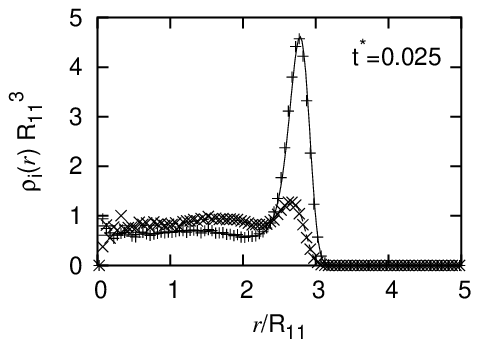}
\end{minipage}
\begin{minipage}[t]{5.1cm}
\includegraphics[width=5cm]{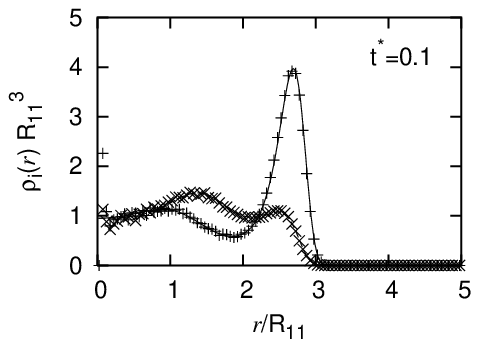}
\end{minipage}
\begin{minipage}[t]{5.1cm}
\includegraphics[width=5cm]{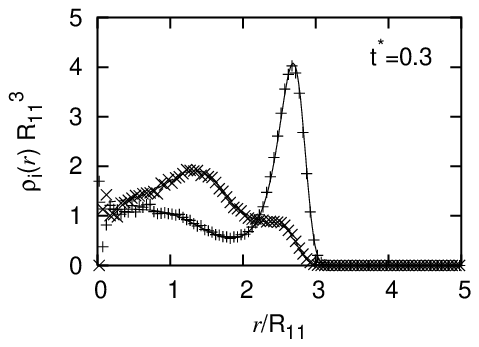}
\end{minipage}
\begin{minipage}[t]{5.1cm}
\includegraphics[width=5cm]{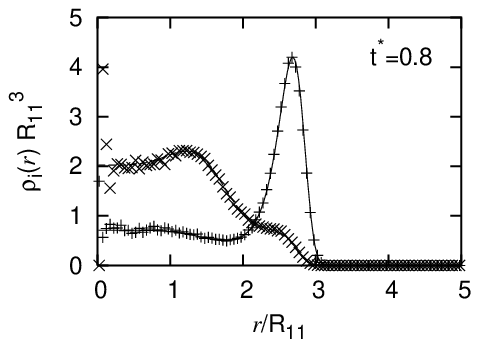}
\end{minipage}
\begin{minipage}[t]{5.1cm}
\includegraphics[width=5cm]{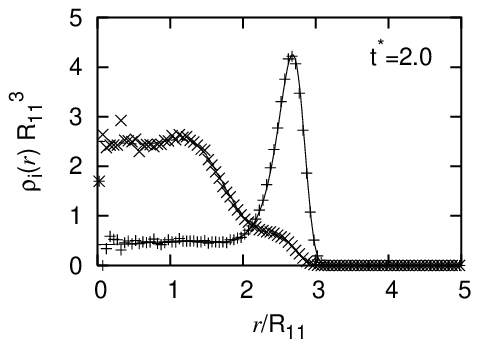}
\end{minipage}
\end{center}
\caption{Density profiles $\rho_i(r,t)$ (DDFT results: solid line for species 1,
dashed line for species 2; symbols are BD results, ($+$) for species 1,
($\times$) for species 2) for a fluid composed of $N_1=200$ particles of species
1 and $N_2=100$ particles
of species 2, which is initially ($t<0$) at equilibrium in an
external potential of the form in Eq.\ (\ref{eq:1}) with ${\cal R}=4R_{11}$.
At $t=0$ the external potentials suddenly change to those
with ${\cal R}=3R_{11}$. The profiles are plotted for various
$t^*=k_BT \Gamma R_{11}^2 t$. The two species of particles are uniformly mixed
in the cavity at $t=0$, but due to the increase in density, the equilibrium
profiles for the fluid in the cavity with ${\cal R}=3R_{11}$ exhibits
a degree of demixing. Note also that $\rho_1(r=0,t)$ is a non-monotonic function
of time.}
\label{fig:1}
\end{figure}

\begin{figure}
\begin{center}
\begin{minipage}[t]{5.1cm}
\includegraphics[width=5cm]{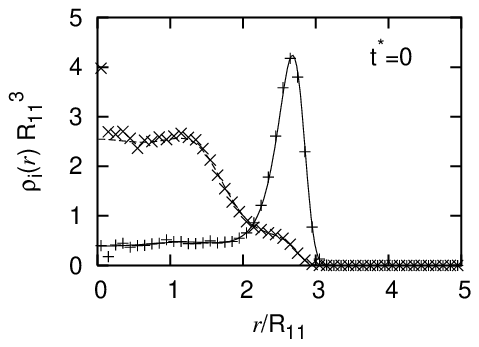}
\end{minipage}
\begin{minipage}[t]{5.1cm}
\includegraphics[width=5cm]{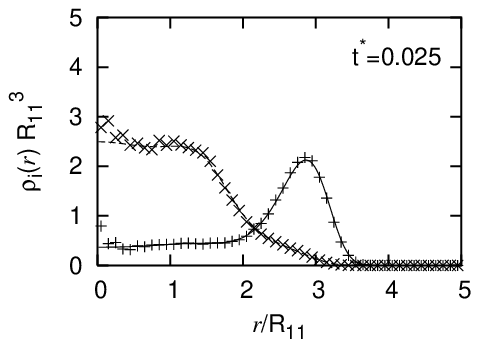}
\end{minipage}
\begin{minipage}[t]{5.1cm}
\includegraphics[width=5cm]{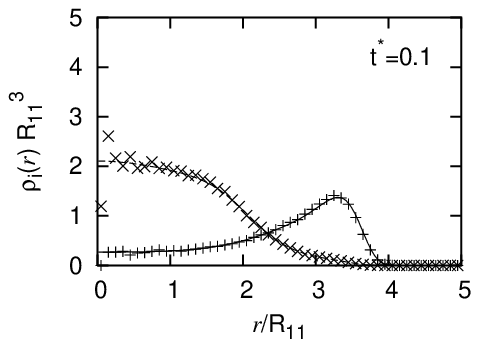}
\end{minipage}
\begin{minipage}[t]{5.1cm}
\includegraphics[width=5cm]{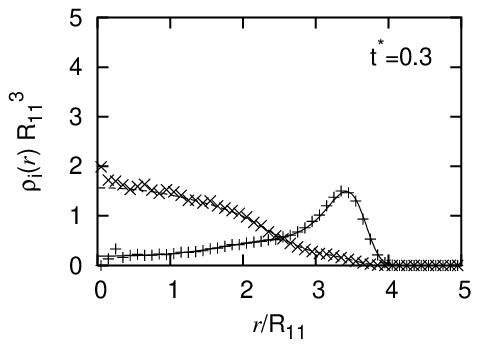}
\end{minipage}
\begin{minipage}[t]{5.1cm}
\includegraphics[width=5cm]{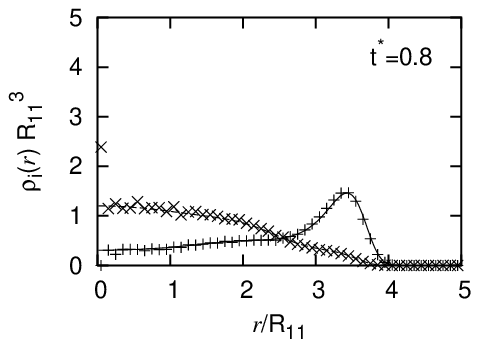}
\end{minipage}
\begin{minipage}[t]{5.1cm}
\includegraphics[width=5cm]{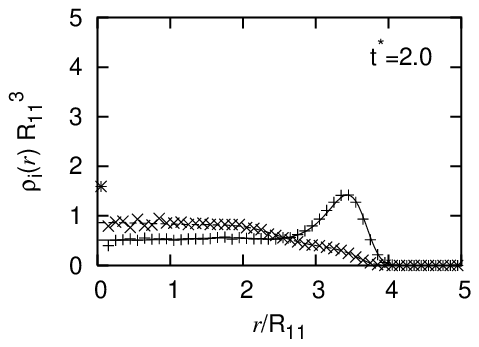}
\end{minipage}
\end{center}
\caption{This is the reverse case of that displayed in Fig.\ \ref{fig:1}.
Initially ($t<0$) the fluid is at equilibrium confined in
external potentials of the form  Eq.\ (\ref{eq:1}) with ${\cal R}=3R_{11}$.
Then at $t=0$ the cavity potentials suddenly change to those
with ${\cal R}=4R_{11}$. The profiles are plotted for various
$t^*=k_BT \Gamma R_{11}^2 t$.}
\label{fig:2}
\end{figure}

\begin{figure}
\includegraphics[width=7.8cm,height=6.2cm]{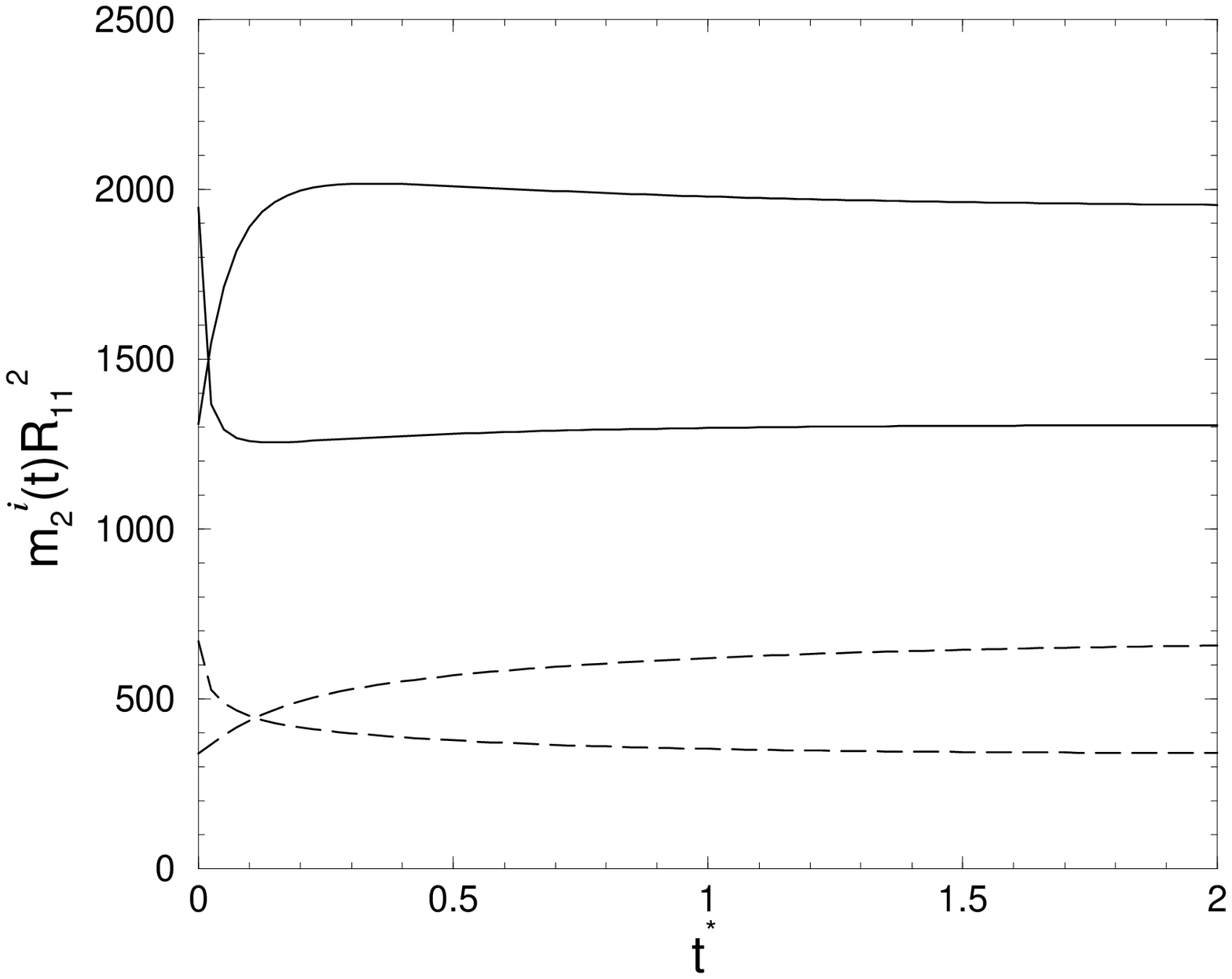}
\includegraphics[width=7.8cm,height=6.2cm]{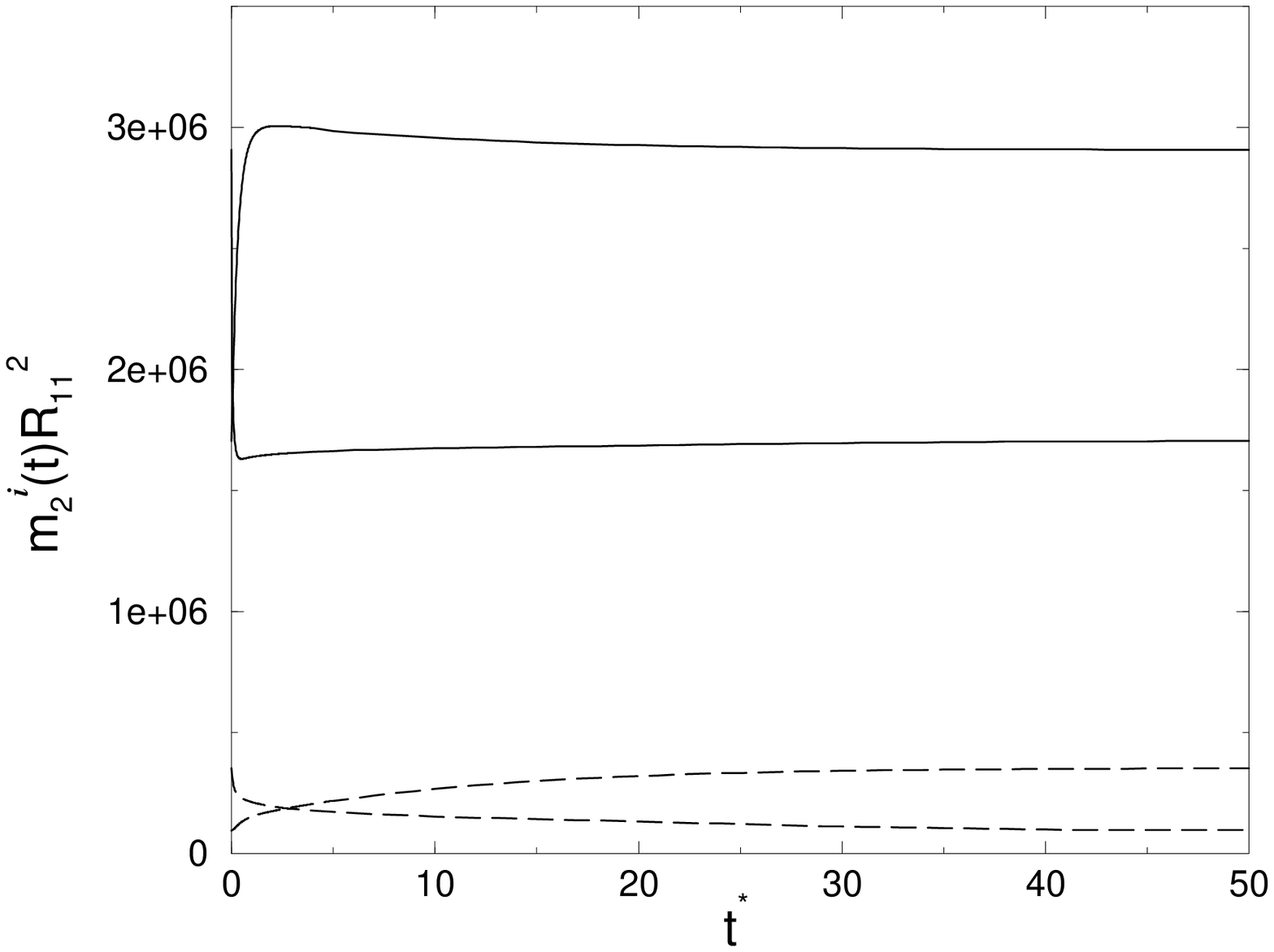}
\caption{The second moment of the density profiles, $m_2^i(t)$, defined by Eq.\
(\ref{eq:2nd_mom}). The left hand figure is for the cases corresponding to
the density
profiles displayed in Figs.\ \ref{fig:1} and \ref{fig:2}. The right hand
figure corresponds to the cases in Figs.\ \ref{fig:4} and \ref{fig:5}. The solid
lines are the second moment for species 1, $m_2^1(t)$, and the dashed lines
$m_2^2(t)$, for species 2.
In each case, the curves with a higher value of $m_2^i(t \rightarrow \infty)$
correspond to a final external potential with a larger value of ${\cal R}$. Note
that in both cases $m_2^1(t)$ is non-monotonic, whereas $m_2^2(t)$ is a
monotonic function of time.}
\label{fig:3}
\end{figure}

\begin{figure}
\begin{center}
\begin{minipage}[t]{5.1cm}
\includegraphics[width=5cm]{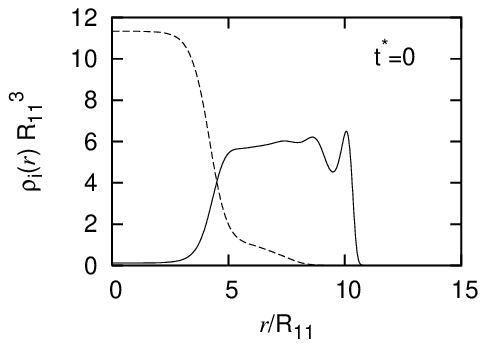}
\end{minipage}
\begin{minipage}[t]{5.1cm}
\includegraphics[width=5cm]{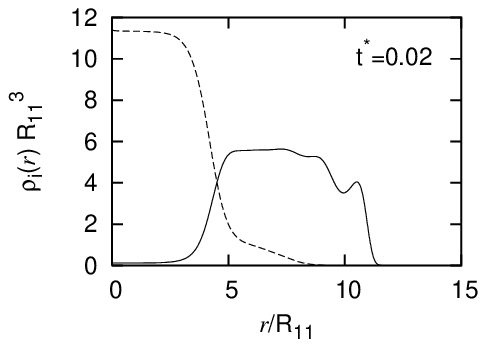}
\end{minipage}
\begin{minipage}[t]{5.1cm}
\includegraphics[width=5cm]{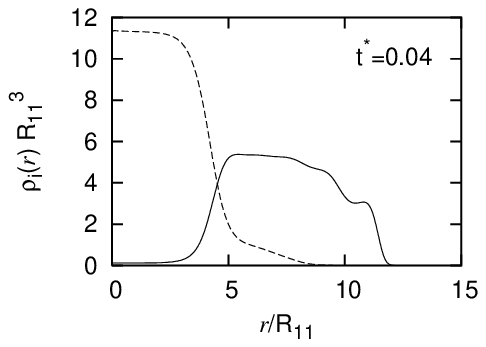}
\end{minipage}
\begin{minipage}[t]{5.1cm}
\includegraphics[width=5cm]{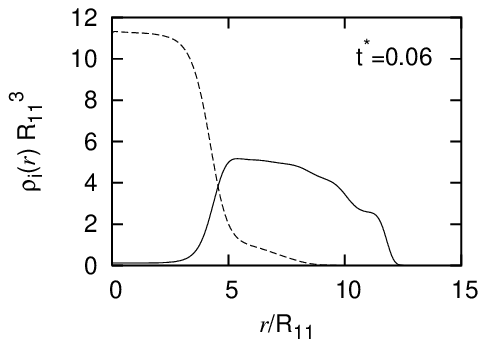}
\end{minipage}
\begin{minipage}[t]{5.1cm}
\includegraphics[width=5cm]{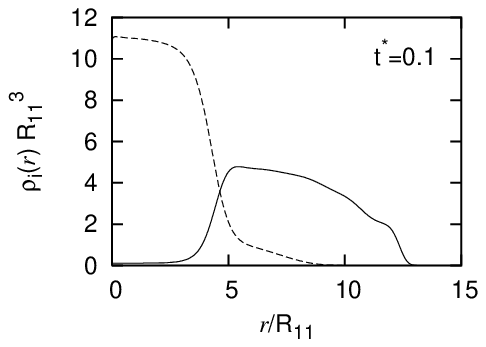}
\end{minipage}
\begin{minipage}[t]{5.1cm}
\includegraphics[width=5cm]{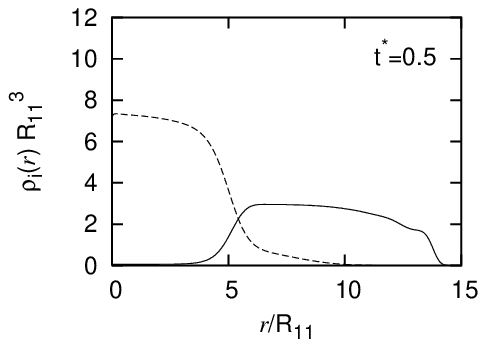}
\end{minipage}
\begin{minipage}[t]{5.1cm}
\includegraphics[width=5cm]{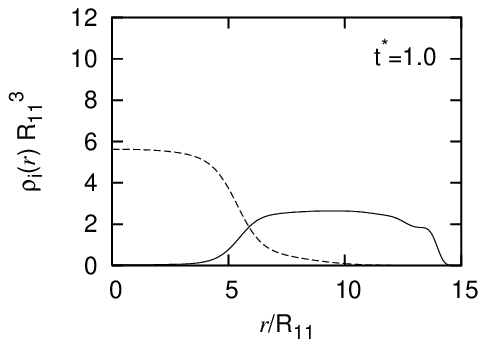}
\end{minipage}
\begin{minipage}[t]{5.1cm}
\includegraphics[width=5cm]{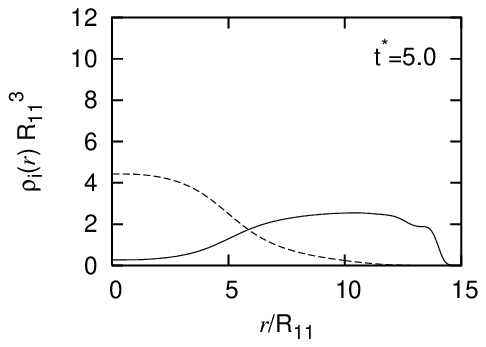}
\end{minipage}
\begin{minipage}[t]{5.1cm}
\includegraphics[width=5cm]{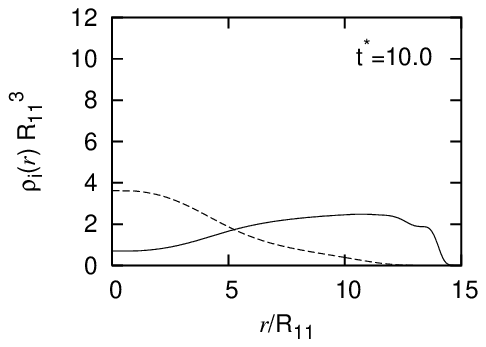}
\end{minipage}
\begin{minipage}[t]{5.1cm}
\includegraphics[width=5cm]{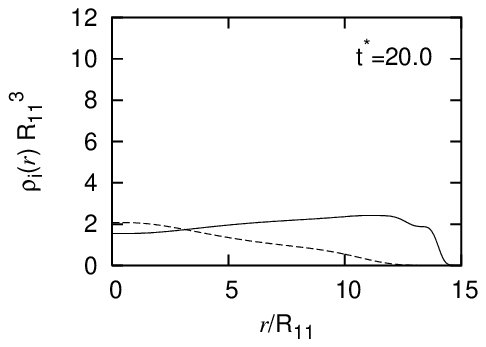}
\end{minipage}
\begin{minipage}[t]{5.1cm}
\includegraphics[width=5cm]{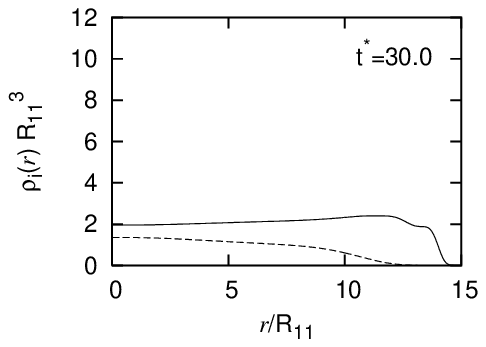}
\end{minipage}
\begin{minipage}[t]{5.1cm}
\includegraphics[width=5cm]{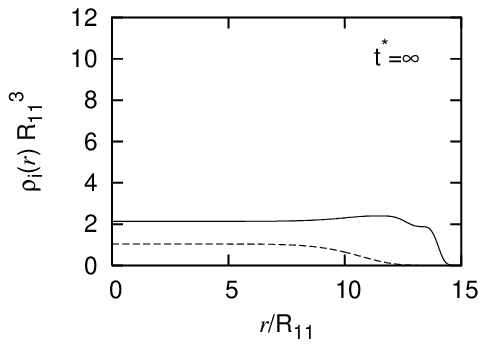}
\end{minipage}
\end{center}
\caption{Density profiles $\rho_i(r,t)$ (solid line for species 1, dashed line
for species 2) for a fluid composed of $N_1=25000$ particles of species 1 and
$N_2=5000$ particles
of species 2, which is initially ($t<0$) at equilibrium confined in
external potentials of the form in Eq.\ (\ref{eq:1}) with ${\cal R}=9R_{11}$.
Then at $t=0$ the external potentials suddenly change to those with
${\cal R}=13R_{11}$. The profiles are plotted for various
$t^*=k_BT \Gamma R_{11}^2 t$. Initially, the fluid is
separated into two phases, one rich in species 1 around the outside, `wetting'
the wall of the cavity, and the other phase in the centre of the cavity, rich
in species 2. Notice the slow diffusion of particles through the fluid-fluid
interface in the later stages of the equilibration.}
\label{fig:4}
\end{figure}

\begin{figure}
\begin{center}
\begin{minipage}[t]{5.1cm}
\includegraphics[width=5cm]{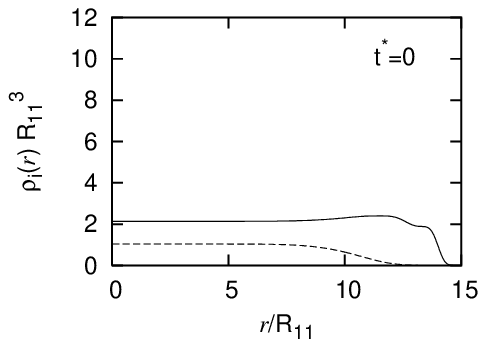}
\end{minipage}
\begin{minipage}[t]{5.1cm}
\includegraphics[width=5cm]{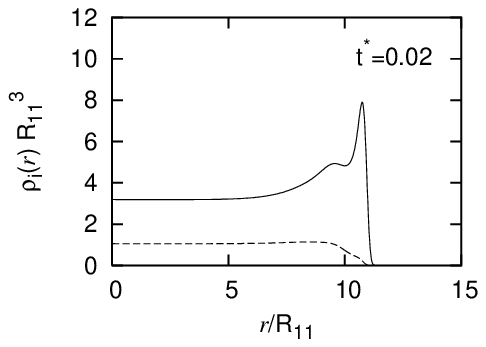}
\end{minipage}
\begin{minipage}[t]{5.1cm}
\includegraphics[width=5cm]{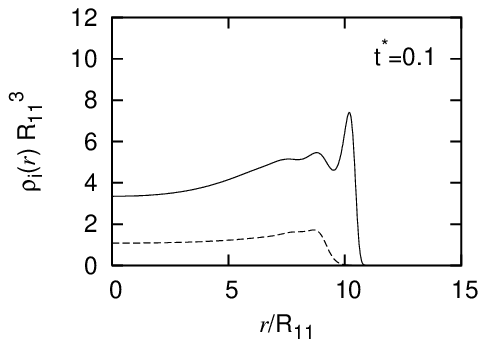}
\end{minipage}
\begin{minipage}[t]{5.1cm}
\includegraphics[width=5cm]{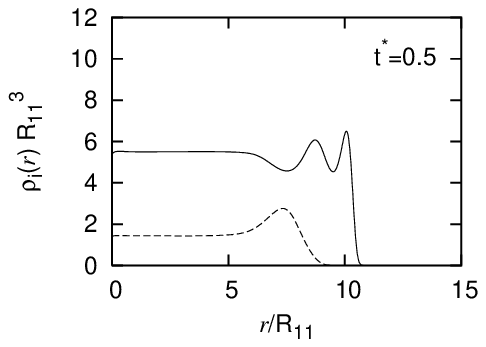}
\end{minipage}
\begin{minipage}[t]{5.1cm}
\includegraphics[width=5cm]{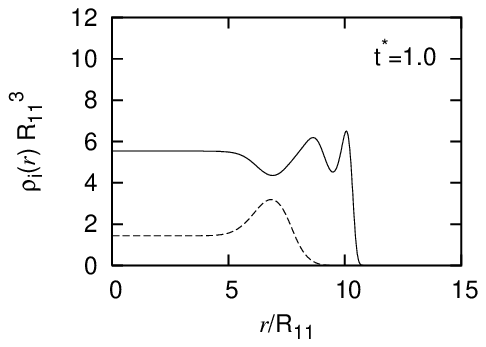}
\end{minipage}
\begin{minipage}[t]{5.1cm}
\includegraphics[width=5cm]{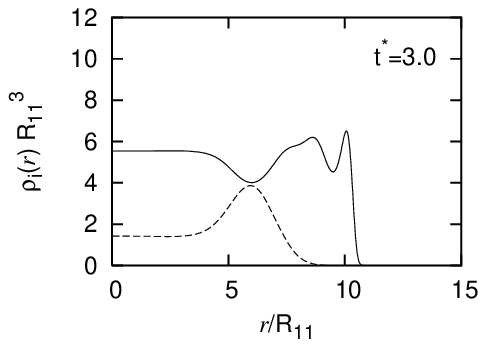}
\end{minipage}
\begin{minipage}[t]{5.1cm}
\includegraphics[width=5cm]{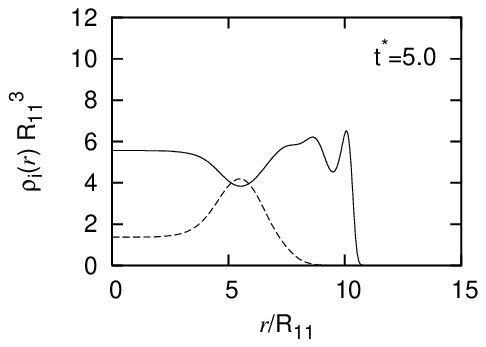}
\end{minipage}
\begin{minipage}[t]{5.1cm}
\includegraphics[width=5cm]{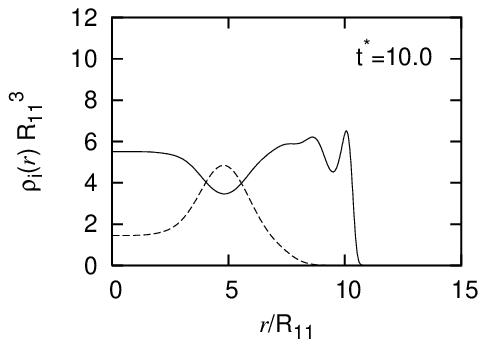}
\end{minipage}
\begin{minipage}[t]{5.1cm}
\includegraphics[width=5cm]{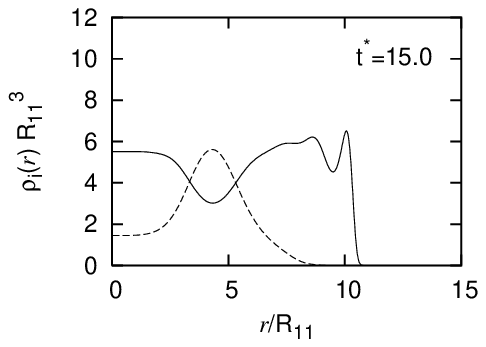}
\end{minipage}
\begin{minipage}[t]{5.1cm}
\includegraphics[width=5cm]{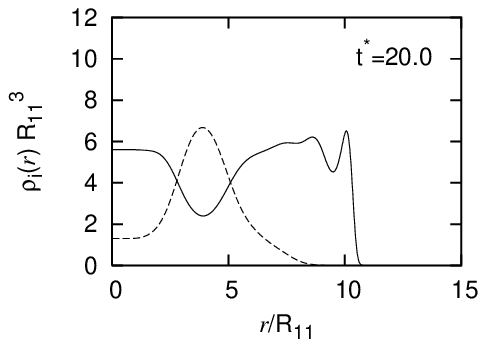}
\end{minipage}
\begin{minipage}[t]{5.1cm}
\includegraphics[width=5cm]{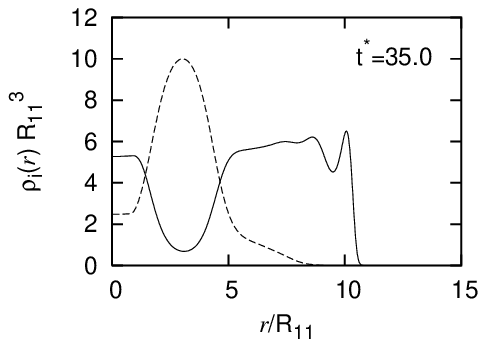}
\end{minipage}
\begin{minipage}[t]{5.1cm}
\includegraphics[width=5cm]{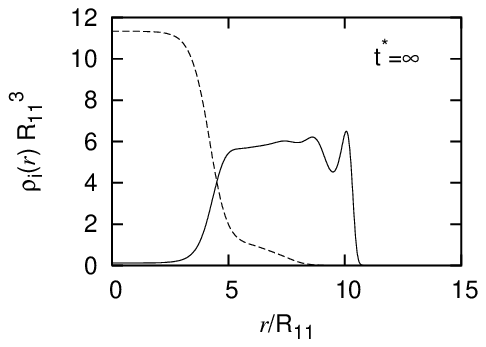}
\end{minipage}
\end{center}
\caption{This is the reverse case of that displayed in Fig.\ \ref{fig:4}.
Initially ($t<0$) the fluid is at equilibrium confined in
external potentials of the form in Eq.\ (\ref{eq:1}) with ${\cal R}=13R_{11}$.
Then at $t=0$ the external potentials suddenly change to
those with ${\cal R}=9R_{11}$. The profiles are plotted for various
$t^*=k_BT \Gamma R_{11}^2 t$. Notice the concentration `wave' (dip
in the profile of species 1 and a peak in profile for species 2) that
travels towards the centre of the cavity which allows the fluid to reach
equilibrium slightly faster than in the opposite (relaxation) process --
see Fig.\ \ref{fig:4}.}
\label{fig:5}
\end{figure}

The first case we consider is that with $N_1=200$ big particles of species 1
and $N_2=100$ particles
of species 2, confined in a cavity with ${\cal R}=4R_{11}$ for $t<0$.
At $t=0$ the cavity size is suddenly reduced to ${\cal R}=3R_{11}$. In Fig.\
\ref{fig:1} we display the evolution of the fluid density profiles after this
sudden compression of the cavity. We display the results from solving the DDFT
(\ref{eq:mainres_multi}) combined with the RPA functional (\ref{eq:F_ex}), as
well as results from BD simulations \cite{AllenTildesley}.
In the BD simulations we numerically integrate Eq.\ (\ref{eq:langevin2})
generalised to two different species of
GCM particles. We typically perform 1500 different runs, each with different
realisations of the stochastic noise term, and then
average over all the different runs in order to obtain the ensemble average
time evolution of the fluid density profiles. In order to generate each starting
configuration, we allow the fluid to equilibrate for a reduced time
$t^* \equiv k_BT \Gamma R_{11}^2 t=10$ with a fixed
parameter ${\cal R}={\cal R}_1(=4R_{11})$ in the external potential, before
changing this parameter to ${\cal R}={\cal R}_2(=3R_{11})$
and determining the relaxation of the
fluid. In Fig.\ \ref{fig:1} the density profiles at $t=0$ correspond to those
of the equilibrium fluid in a cavity with ${\cal R}=4R_{11}$\footnote{
Note that within the grand canonical EDFT the {\em mean} number of
particles $\left< N_i \right>$ of species $i$ in the cavity is constrained.
In practice we set the chemical potentials $\mu_i$ such that the the
mean number of particles in the cavity, $\left< N_i \right>=\int \dr \rho_i(\rr)
=N_i$, the number of particles in the BD simulation.}.
When the external potential parameter ${\cal R}=4R_{11}$,
the two species of GCM particles are mixed within the cavity,
although because the walls of the cavity favour species 1, there is a higher
density of species 1 around the outside of the cavity. This preference of the
cavity wall for species 1 is due to the fact that the cavity wall
potentials for both
species of particles decay into the fluid with the same decay length. This
results in an effective attraction between the wall and species 1
\cite{Archer2}. For small $t>0$, the initial `compression' of the fluid results
in the growth of a sharp peak in the densities of both species around $r \simeq
{\cal R}_2=3R_{11}$. This causes a density `wave' to travel through the fluid
into the centre of the cavity. The fluid reaches equilibrium at
$t^* \equiv t/\tau_B \simeq 1.3$, where $\tau_B=\beta/\Gamma R_{11}^2$ is the
Brownian time scale. In Fig.\ \ref{fig:1} we display the fluid
density profiles for the times $t^*=0$, 0.025, 0.1, 0.3, 0.8, and 2. The profile
for $t^*=2$ effectively corresponds to the equilibrium profile for the external
potential (\ref{eq:1}) with ${\cal R}=3R_{11}$. With this value of ${\cal R}$
the fluid exhibits a degree of phase separation in the cavity due to the fact
that the total
density of the fluid in the cavity has increased. The particles of species 1 are
mostly to be found adsorbed around the wall of the cavity and species 2 towards
the centre of the cavity. In Fig.\ \ref{fig:2} we display the density profiles
for the reverse situation to the case in Fig.\ \ref{fig:1}: The external
potentials (\ref{eq:1}) are initially ($t<0$) those with ${\cal R}=3R_{11}$.
Then, at $t=0$, the potentials are changed to those with ${\cal R}=4R_{11}$.
In this case the
fluid reaches equilibrium when $t^* \simeq 1.7$. The reason that the
`compression' case (Fig.\ \ref{fig:1}) is able to reach equilibrium faster than
the relaxation case (Fig.\ \ref{fig:2}) is that in the compression case the
`shock wave' that travels inward, mixes up the fluid, which
allows it to reach equilibrium faster than in the relaxation case.
In Fig.\ \ref{fig:3} we display the second moment of the
density profile,
\begin{equation}
m_2^i(t) = \int \dr r^2 \rho_i(\rr,t).
\label{eq:2nd_mom}
\end{equation}
Dzubiella and Likos \cite{joe:christos} found that for a one component GCM fluid
this is a monotonic function of time. For the relaxation (cavity expansion)
case they found that
$m_2(t)$ can be very accurately approximated by an exponentially decaying
function. For the compression case, $m_2(t)$ is also monotonic, and can be
accurately parameterised by a function composed of two exponentials. They also
found that the one component GCM fluid reaches equilibrium faster in the
compression case,
where ${\cal R}$ is decreased at time $t=0$, rather than the relaxation case.
For the binary GCM fluid, $m_2^2(t)$ is also a monotonic function of time
(dashed lines), but
interestingly, $m_2^1(t)$ is not (solid lines) -- see Fig.\
\ref{fig:3}. This is because particles of species 1 are
adsorbed on the wall of the cavity and when the parameter
${\cal R}$ in the external potentials is suddenly decreased,
particles of species 1 are forced more strongly
towards the centre of the cavity than the species 2 particles.
This causes an increase in the
density of species 1 at the centre of the cavity. However, in the final stages
of the fluid equilibration, the density of species 1 at the centre of the cavity
decreases again. In other words, $\rho_1(r=0,t)$ is a non-monotonic function of
time. Similarly, when the cavity is suddenly increased in size, it is the
particles of species 1 that `feel' the space around the outside of the fluid
that has suddenly appeared, rather than the species 2 particles,
and so it is species 1 particles that move outwards to fill
this space. However, in the final stages of the fluid equilibration process,
there is a net flow of species 1 particles back towards the centre of the
cavity. For these reasons, only $m_2^1(t)$ is a non-monotonic function of time.
Dynamic processes such as these only occur when the fluid confined in
the cavity is a binary fluid, where the wall of the cavity has a preference for
one of the species. As can be seen from Figs.\ \ref{fig:1} and \ref{fig:2} there
is remarkably good agreement between the density profiles obtained from the DDFT
and from the BD simulations. This agreement gives us confidence concerning the
reliability of the DDFT for cases where the numbers of particles are such that
BD simulations become computationally too expensive. Note also that the RPA
functional (\ref{eq:F_ex}) becomes increasingly accurate as the GCM
fluid density is increased \cite{Likos,LangJPCM}; this should further
improve the accuracy of the DDFT results.

In Figs.\ \ref{fig:4} and \ref{fig:5} we display the DDFT results for a similar
situation as in Figs.\ \ref{fig:1} and \ref{fig:2},
except in this case the cavity is larger and the number of
confined particles is higher: $N_1=25000$ particles of species 1 and
$N_2=5000$ particles
of species 2 (the average densities are also higher).
In Fig.\ \ref{fig:4} we display the results for the
case when the fluid, for $t<0$, is at equilibrium in a cavity with potentials
given by Eq.\
(\ref{eq:1}) with ${\cal R}=9R_{11}$. Then at $t=0$ the potentials change
suddenly to those with ${\cal R}=13R_{11}$. We plot the density profiles for 
$t^*=0$, 0.02, 0.04, 0.06, 0.1, 0.5, 1, 5, 10, 20, 30 and $\infty$. (Here and
elsewhere, the $t=\infty$ density profiles are those obtained using EDFT for
cavity potentials with ${\cal R}={\cal R}_2$. In the present case
${\cal R}_2=13R_{11}$.) We see that
at $t=0$ the fluid is strongly phase separated in the cavity, with the phase
rich in species 1 `wetting' the wall of the cavity, and the phase rich in
species 2 at the centre of the cavity. The final equilibrium configuration,
$t \rightarrow \infty$, when the cavity radius ${\cal R}=13R_{11}$,
is that where the two species of particles are mixed
together in the cavity, although the preference of the cavity wall for
species 1 ensures that
the density of species 1 is still higher around the outside of the
cavity. In order to reach this equilibrium configuration the fluid first
exhibits a `quick' flow of the phase rich in species 1 to fill the space created
around the outside of the fluid by the cavity expansion. There is then a second
`slow' process whereby the two demixed phases mix, i.e.\ the diffusion of
particles across the fluid-fluid interface is a slow process.

For the reverse (cavity compression) situation, the results are displayed in
Fig.\ \ref{fig:5}.
For $t<0$ the fluid is at equilibrium in a cavity with potentials given by Eq.\
(\ref{eq:1}) with ${\cal R}=13R_{11}$. Then at $t=0$ the potentials change
suddenly to those with ${\cal R}=9R_{11}$. We plot the density profiles for 
$t^*=0$, 0.02, 0.1, 0.5, 1, 3, 5, 10, 15, 20, 35 and $\infty$. In this case the
fluid reaches equilibrium slightly faster than in the case where the cavity
size is increased.
This is because in the cavity compression case, the sudden decrease of
${\cal R}$ at $t=0$ sends a particle concentration `wave' into the centre of the
cavity (see Fig.\ \ref{fig:5}) which allows the fluid to reach equilibrium
faster. This `wave' forms as a
dip in the density profile of species 1 and a peak in profile for species 2 
couple together and move towards the centre of the cavity. The amplitude of the
wave increases with proximity to the centre of the cavity.
As can be seen in the right hand figure of Fig.\ \ref{fig:3} for both
these cases the second moments of the density profiles, given by Eq.\
(\ref{eq:2nd_mom}), show similar behaviour to the cases with fewer particles in
Figs.\ \ref{fig:1} and \ref{fig:2} -- see Fig.\ \ref{fig:3}. $m_2^1(t)$ is a
non-monotonic function of time, whereas $m_2^2(t)$ is a monotonic function of
time. As for the cases in Figs.\ \ref{fig:1} and \ref{fig:2}, this is due to the
fact that the species 1 particles are adsorbed on the cavity wall, rather than
the species 2 particles.

\subsection{DDFT for a GCM fluid which exhibits microphase separation} 

\begin{figure}
\begin{center}
\begin{minipage}[t]{5.1cm}
\includegraphics[width=5cm]{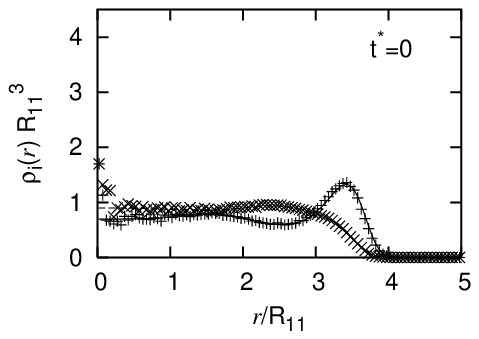}
\end{minipage}
\begin{minipage}[t]{5.1cm}
\includegraphics[width=5cm]{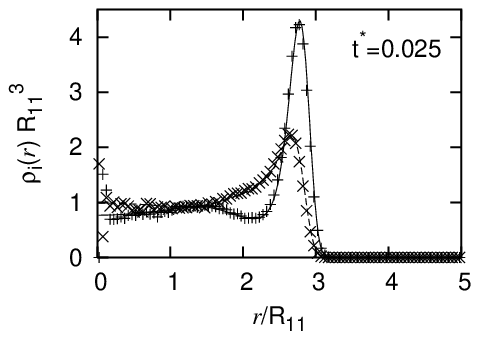}
\end{minipage}
\begin{minipage}[t]{5.1cm}
\includegraphics[width=5cm]{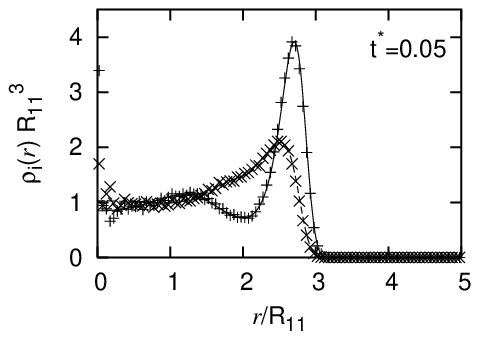}
\end{minipage}
\begin{minipage}[t]{5.1cm}
\includegraphics[width=5cm]{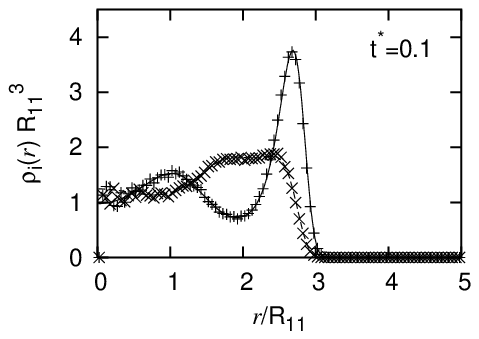}
\end{minipage}
\begin{minipage}[t]{5.1cm}
\includegraphics[width=5cm]{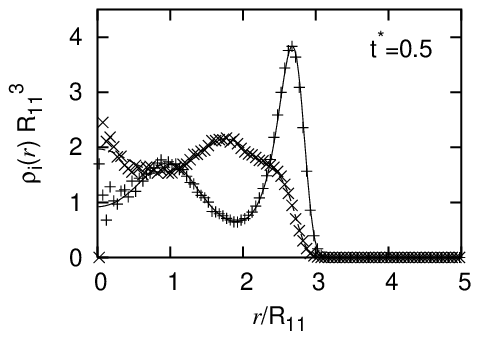}
\end{minipage}
\begin{minipage}[t]{5.1cm}
\includegraphics[width=5cm]{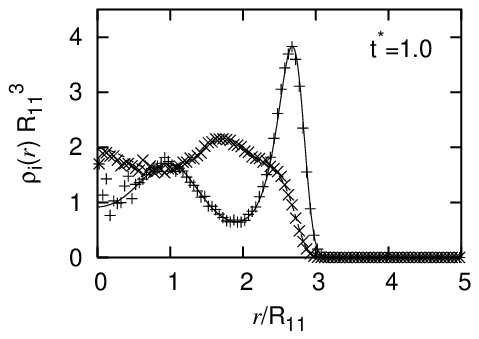}
\end{minipage}
\end{center}
\caption{Density profiles $\rho_i(r,t)$ (solid line is DDFT results
for species 1, dashed line for species 2;
symbols are BD results, ($+$) for species 1, ($\times$)
for species 2.) for a fluid composed of $N_1=200$ particles of species 1 and
$N_2=150$ particles
of species 2, which is initially ($t<0$) at equilibrium confined in
external potentials of the form in Eq.\ (\ref{eq:1}) with ${\cal R}=4R_{11}$.
Then at $t=0$, the external potentials suddenly change to those
with ${\cal R}=3R_{11}$. The profiles are
plotted for various $t^*=k_BT \Gamma R_{11}^2 t$. This model fluid exhibits
microphase-separation. The final $t^*=1$ configuration is an `onion' structure.}
\label{fig:6}
\end{figure}

\begin{figure}
\begin{center}
\begin{minipage}[t]{5.1cm}
\includegraphics[width=5cm]{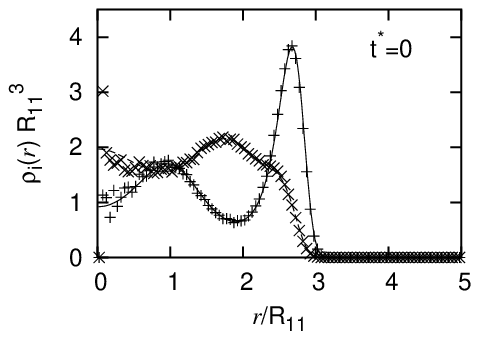}
\end{minipage}
\begin{minipage}[t]{5.1cm}
\includegraphics[width=5cm]{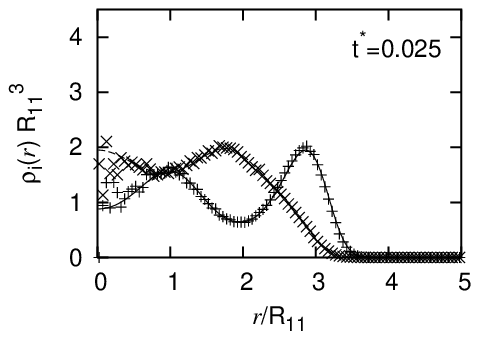}
\end{minipage}
\begin{minipage}[t]{5.1cm}
\includegraphics[width=5cm]{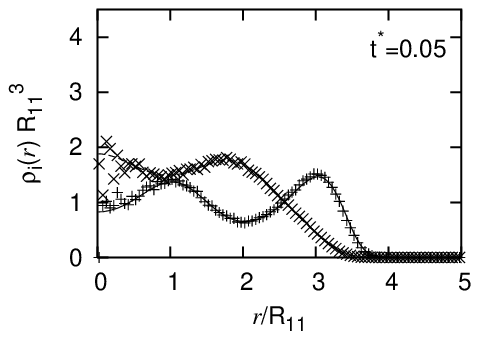}
\end{minipage}
\begin{minipage}[t]{5.1cm}
\includegraphics[width=5cm]{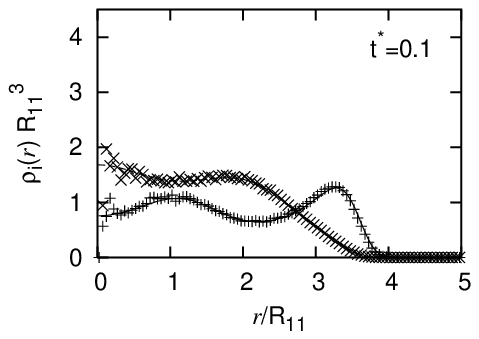}
\end{minipage}
\begin{minipage}[t]{5.1cm}
\includegraphics[width=5cm]{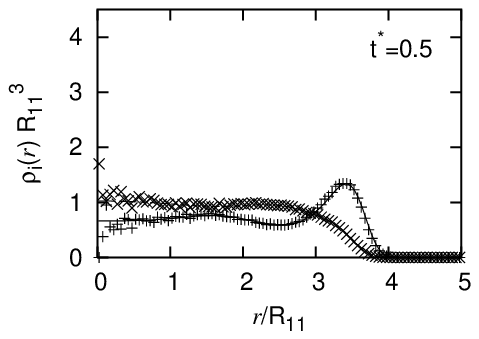}
\end{minipage}
\begin{minipage}[t]{5.1cm}
\includegraphics[width=5cm]{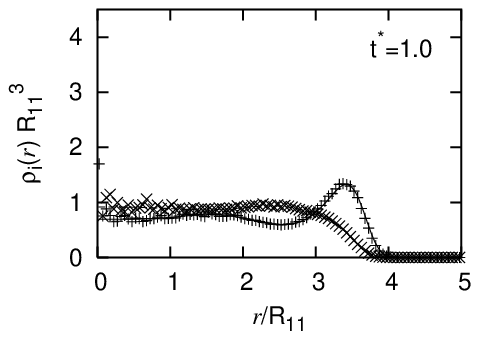}
\end{minipage}
\end{center}
\caption{This is the reverse case of that displayed in Fig.\ \ref{fig:6}.
Initially ($t<0$) the fluid is at equilibrium confined in
external potentials of the form in Eq.\ (\ref{eq:1}) with ${\cal R}=3R_{11}$.
Then at $t=0$, the external potentials suddenly change
to those with ${\cal R}=4R_{11}$. The profiles are
plotted for various $t^*=k_BT \Gamma R_{11}^2 t$. The initial $t=0$
configuration is an `onion' structure.}
\label{fig:7}
\end{figure}

\begin{figure}
\begin{center}
\begin{minipage}[t]{5.1cm}
\includegraphics[width=5cm]{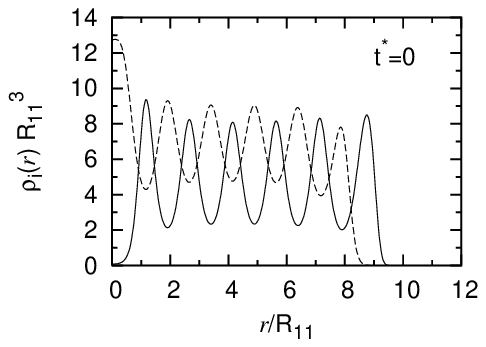}
\end{minipage}
\begin{minipage}[t]{5.1cm}
\includegraphics[width=5cm]{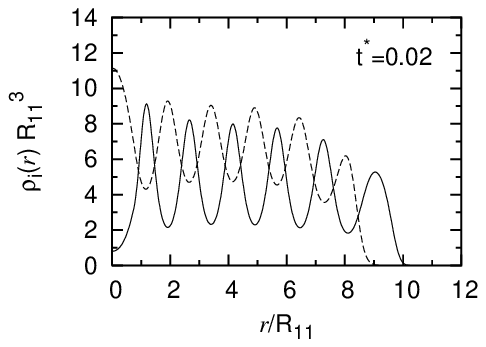}
\end{minipage}
\begin{minipage}[t]{5.1cm}
\includegraphics[width=5cm]{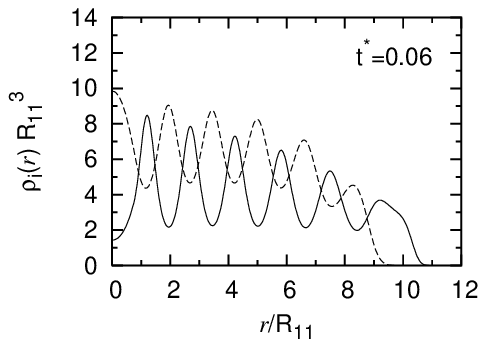}
\end{minipage}
\begin{minipage}[t]{5.1cm}
\includegraphics[width=5cm]{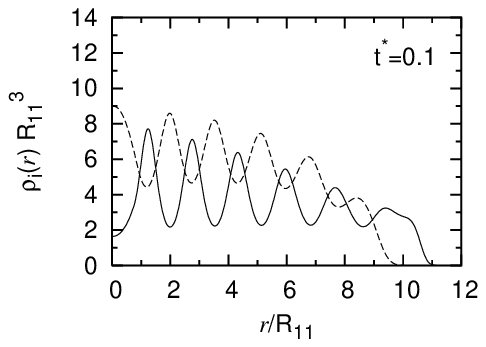}
\end{minipage}
\begin{minipage}[t]{5.1cm}
\includegraphics[width=5cm]{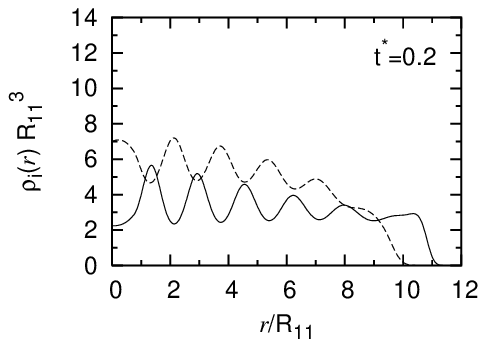}
\end{minipage}
\begin{minipage}[t]{5.1cm}
\includegraphics[width=5cm]{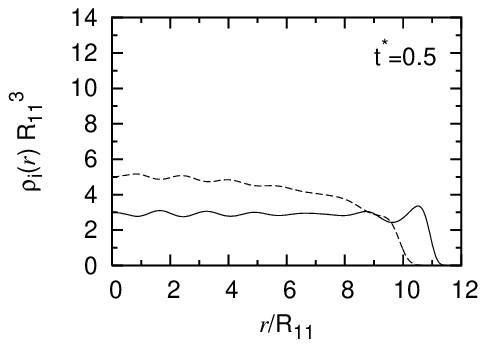}
\end{minipage}
\begin{minipage}[t]{5.1cm}
\includegraphics[width=5cm]{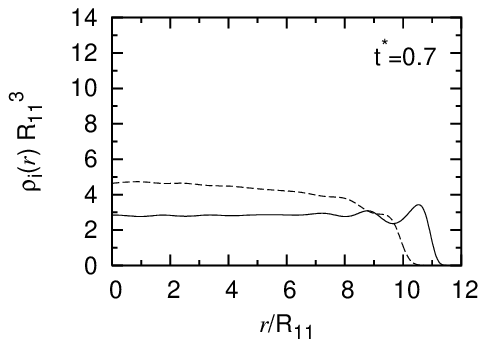}
\end{minipage}
\begin{minipage}[t]{5.1cm}
\includegraphics[width=5cm]{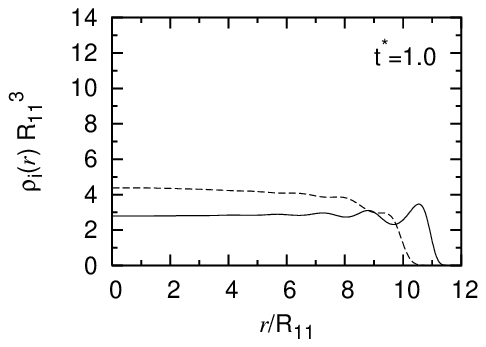}
\end{minipage}
\begin{minipage}[t]{5.1cm}
\includegraphics[width=5cm]{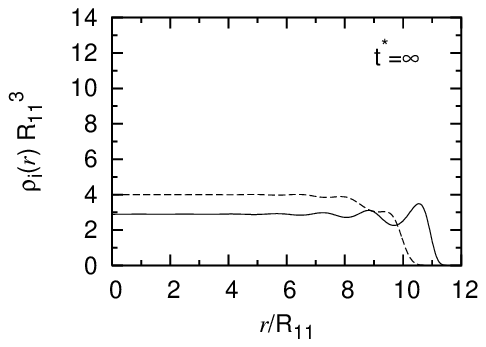}
\end{minipage}
\end{center}
\caption{Density profiles $\rho_i(r,t)$ (solid line for species 1, dashed line
for species 2) for a fluid composed of $N_1=16000$ particles of species 1 and
$N_2=15000$ particles
of species 2, which is initially ($t<0$) is at equilibrium confined
in external potentials of the form in Eq.\ (\ref{eq:1}) with ${\cal R}=8R_{11}$.
Then at $t=0$ the external potentials are suddenly changed to those
with ${\cal R}=10R_{11}$. The profiles are plotted for various
$t^*=k_BT \Gamma R_{11}^2 t$. This model fluid exhibits
microphase-separation. The initial $t=0$ configuration is an `onion' structure.}
\label{fig:9}
\end{figure}

\begin{figure}
\begin{center}
\begin{minipage}[t]{5.1cm}
\includegraphics[width=5cm]{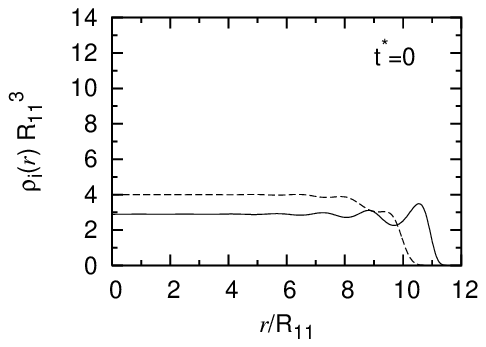}
\end{minipage}
\begin{minipage}[t]{5.1cm}
\includegraphics[width=5cm]{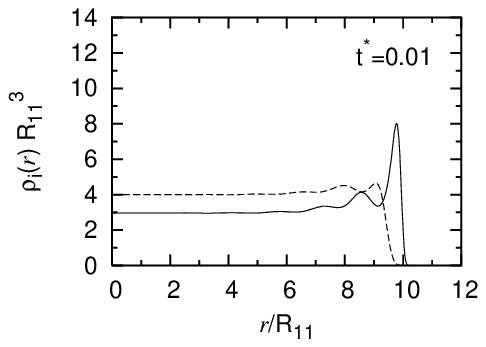}
\end{minipage}
\begin{minipage}[t]{5.1cm}
\includegraphics[width=5cm]{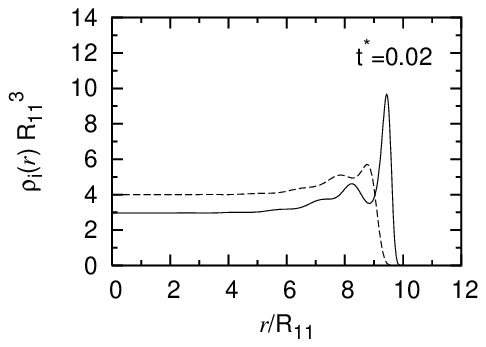}
\end{minipage}
\begin{minipage}[t]{5.1cm}
\includegraphics[width=5cm]{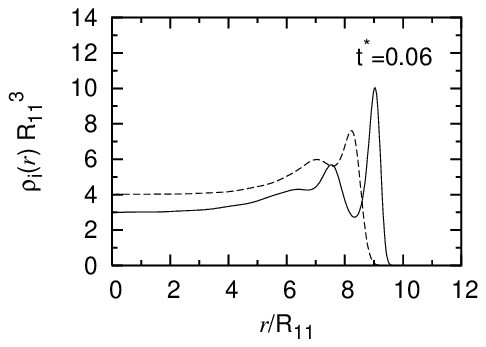}
\end{minipage}
\begin{minipage}[t]{5.1cm}
\includegraphics[width=5cm]{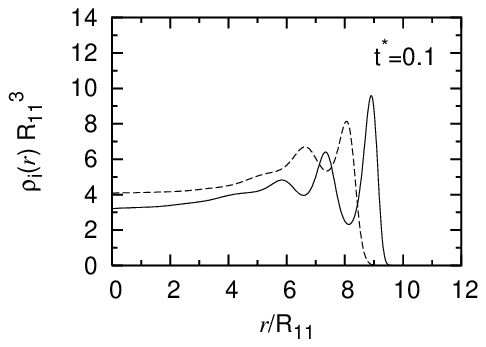}
\end{minipage}
\begin{minipage}[t]{5.1cm}
\includegraphics[width=5cm]{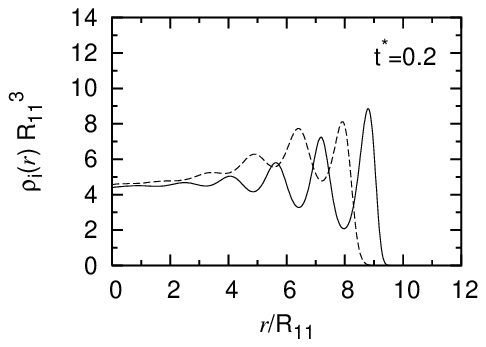}
\end{minipage}
\begin{minipage}[t]{5.1cm}
\includegraphics[width=5cm]{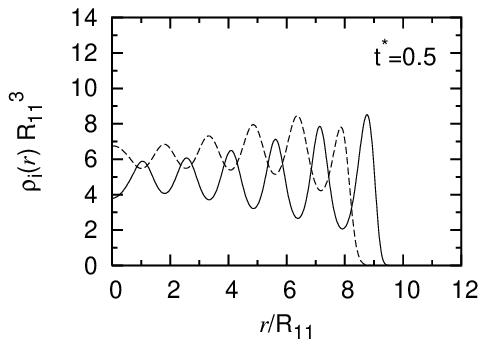}
\end{minipage}
\begin{minipage}[t]{5.1cm}
\includegraphics[width=5cm]{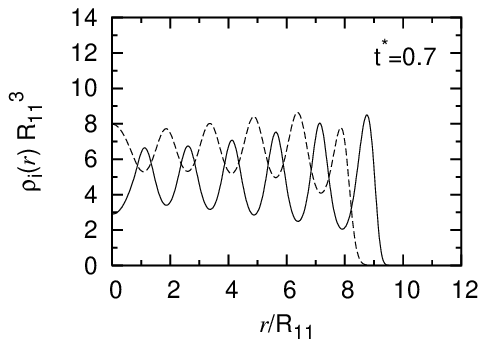}
\end{minipage}
\begin{minipage}[t]{5.1cm}
\includegraphics[width=5cm]{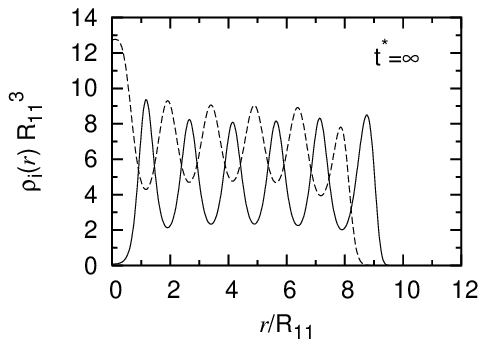}
\end{minipage}
\end{center}
\caption{This is the reverse case of that displayed in Fig.\ \ref{fig:9}.
Initially ($t<0$) the fluid is at equilibrium confined in
external potentials of the form in Eq.\ (\ref{eq:1}) with ${\cal R}=10R_{11}$.
Then at $t=0$, the external potentials suddenly change to
those with ${\cal R}=8R_{11}$. The density profiles (solid line
for species 1, dashed line for species 2) are plotted for various
$t^*=k_BT \Gamma R_{11}^2 t$.}
\label{fig:10}
\end{figure}

We now consider a binary GCM fluid with pair potential parameters the same as in
the previous subsection except that $R_{12}=0.6R_{11}$, i.e.\
$R_{12}=(1+\Delta)(R_{11}+R_{22})/2$, with $\Delta<0$; the fluid
exhibits negative non-additivity. This fluid does not exhibit bulk fluid-fluid
phase separation. Due to the negative non-additivity there is a propensity
to ordering in the fluid in which particles of species 1 have as nearest
neighbours particles of species 2, and vice versa \cite{Archer6}. We refer to
this phenomenon as `1-2 ordering'. At high total densities EDFT predicts that
this GCM fluid freezes into a crystal in which the particles are highly
delocalized, with Lindemann parameters as high as 90\%
near melting \cite{Archer6}. When the fluid is confined in a spherical cavity
and the fluid density is high enough,
the fluid forms an `onion' structure of alternating layers, one
particle thick, of the two different species \cite{Archer6} (see also Figs.\
\ref{fig:6}, \ref{fig:7}, \ref{fig:9} and \ref{fig:10}).
Similarly, when the high density mixture is confined in a planar slit the
density profiles show that fluid forms alternating layers of the two different
species parallel to the walls of the slit \cite{Archer6}.
At lower densities the
two different species of particles are mixed. This microphase separation is
associated with the fact that the bulk
fluid exhibits an instability with respect to periodic
density modulations -- a `$\lambda$-instability'. For more details regarding
the origin of this instability see Ref.\ \cite{Archer6} and references therein.
Here our interest is limited to the question: Can the DDFT
describe the formation of `onion' structures in a spherical cavity with
potentials given by Eq.\ (\ref{eq:1}) when ${\cal R}$ is reduced, and can the
DDFT describe the onion `melting' when ${\cal R}$ is increased?

\begin{figure}
\begin{center}
\includegraphics[width=7.8cm]{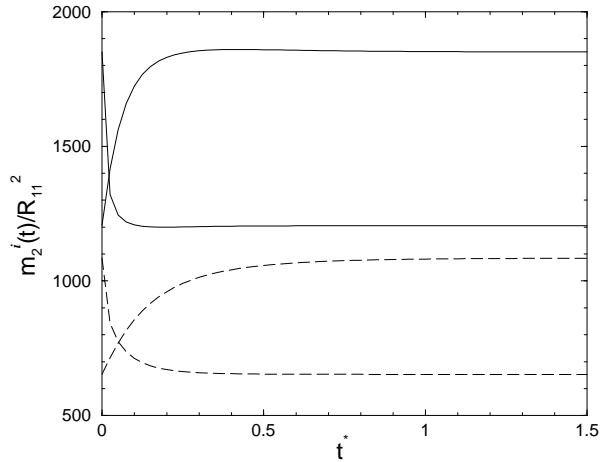}
\end{center}
\caption{The second moment of the density profiles, $m_2^i(t)$, defined by Eq.\
(\ref{eq:2nd_mom}), for a GCM fluid exhibiting
microphase-separation, with density
profiles displayed in Figs.\ \ref{fig:6} and \ref{fig:7}.
The solid lines are the second moment for species 1, $m_2^1(t)$, and the dashed
lines $m_2^2(t)$, for species 2.
In each case, the curves with a higher value of $m_2^i(t \rightarrow \infty)$
correspond to a final external potential with a larger value of ${\cal R}$.
Note that in both cases $m_2^1(t)$ is non-monotonic, whereas $m_2^2(t)$ is a
monotonic function of time, as was the case in Fig.\ \ref{fig:3} for a GCM fluid
that exhibits bulk phase separation.}
\label{fig:8}
\end{figure}

We follow a strategy similar to that taken in the previous subsection and make a
comparison between the results from DDFT and BD simulations for cases with a
limited number of particles, in order to demonstrate the reliability of the DDFT
for this particular GCM fluid and then we
apply the DDFT to cases where the number of particles is too high for
simulations to be a realistic tool for studying the dynamics.
In Fig.\ \ref{fig:6} we display the
density profiles for a fluid composed of $N_1=200$ particles of species 1 and
$N_2=150$ particles
of species 2, which is initially ($t<0$) at equilibrium confined in
external potentials of the form in Eq.\ (\ref{eq:1}) with ${\cal R}=4R_{11}$.
Then at $t=0$ the external potentials are suddenly changed to
those with ${\cal R}=3R_{11}$. We display the density
profiles for $t^*=0$, 0.025, 0.05, 0.1, 0.5 and 1. In the initial $t=0$ density
profiles we see that there is little sign of the fluid exhibiting 1-2 ordering.
As was the case for the fluid described in the previous subsection, the wall
potentials result in an effective attraction between the wall of the cavity and
particles of
species 1. Therefore there is an increased density of species 1 around the wall
of the cavity. Following the compression in the cavity at $t=0$, there is an
initial in-flow of fluid towards the centre of the cavity and an
increase in the density of the fluid near to the cavity
wall. Following this initial stage, the fluid relaxes into a configuration which
exhibits a pronounced degree of 1-2 ordering -- see the $t^*=1$ figure in Fig.\
\ref{fig:6}. In Fig.\ \ref{fig:7} we display the results for the reverse
situation, i.e.\ when the fluid is initially ($t<0$) at equilibrium confined in
external potentials of the form in Eq.\ (\ref{eq:1}) with ${\cal R}=3R_{11}$,
then at $t=0$, the external potentials suddenly change to those
with ${\cal R}=4R_{11}$. We display the
profiles for $t^*=0$, 0.025, 0.05, 0.1, 0.5 and 1. In both cases the agreement
between the DDFT and the BD simulations is remarkably good. Fig.\ \ref{fig:8}
displays the results for the second moment of the density profile, defined by
Eq.\ (\ref{eq:2nd_mom}). As in the previous subsection, in the present case the
moments $m_2^1(t)$ for species 1 are non-monotonic functions of time, whereas
the moments $m_2^2(t)$ are monotonic functions of time.
The fluid which exhibits 1-2 ordering is able to reach equilibrium faster
than the fluid considered in the previous
subsection which exhibits bulk phase separation. This is because the average
distance the particles must diffuse to be arranged in a microphase-separated
distribution is shorter than in the case where the final equilibrium
configuration is that exhibiting `bulk' phase-separation. In the latter case the
particles of species 1 must diffuse to the outside of the cavity, whilst
particles of species 2 must diffuse to the centre of the cavity in order
to reach equilibrium.

In Figs.\ \ref{fig:9} and \ref{fig:10} we display the DDFT results for a similar
situation, except now the cavity is larger and the number of
confined particles is also much bigger: $N_1=16000$ particles of species 1 and
$N_2=15000$ particles
of species 2 (the average densities are also higher).
In Fig.\ \ref{fig:9} we display the results for the
case when the fluid, for $t<0$, is at equilibrium in a cavity with potentials
given by Eq.\ (\ref{eq:1}) with ${\cal R}=8R_{11}$. Then at $t=0$ the potentials
change suddenly to those with ${\cal R}=10R_{11}$. We plot the density profiles
for  $t^*=0$, 0.02, 0.06, 0.1, 0.2, 0.5, 0.7, 1 and $\infty$.
At $t=0$ the two different species of particles are strongly ordered into
alternating layers of the two different species -- an `onion' structure.
The final equilibrium configuration,
$t \rightarrow \infty$, is that where the two species of particles are uniformly
mixed in the cavity, although, due to the preference of the cavity wall for
species 1, the density of species 1 is higher in the outer region of the
cavity. The equilibration process is almost an order of magnitude shorter in
time than
in either of the cases in Figs.\ \ref{fig:4} or \ref{fig:5} for the GCM fluid
exhibiting bulk liquid-liquid phase separation, even though the number of
particles involved in these cases is similar. This is because the
particles must diffuse a distance $\sim R_{11}$, the particle size,
in order to go from a state with
1-2 ordering to a mixed state, whereas for the fluid exhibiting `bulk' phase
separation the particles must diffuse a much larger distance $\sim {\cal R}$,
the radius of the cavity.
For the reverse (cavity compression) situation, the results are displayed in
Fig.\ \ref{fig:10}. Initially, for $t<0$,
the fluid is at equilibrium in a cavity with potentials given by Eq.\
(\ref{eq:1}) with ${\cal R}=10R_{11}$. Then at $t=0$ the potentials change
suddenly to those with ${\cal R}=8R_{11}$. We plot the density profiles for 
$t^*=0$, 0.01, 0.02, 0.06, 0.1, 0.2, 0.5, 0.7 and $\infty$. As with the cavity
expansion case in Fig.\ \ref{fig:9} this process is almost an order of magnitude
quicker in time
than in either of the cases in Figs.\ \ref{fig:4} or \ref{fig:5}. The
compression causes the total fluid density to first increase around the wall of
the cavity, before increasing near the centre of the cavity. This results in the
`onion' layers forming first in the outer region and then developing inwards
towards the centre of the cavity as the fluid equilibrates.

\section{Liouville, Kramers and Smoluchowski equations}
\label{sec:LKS}

We now turn to the question of what approximations are involved in the present
description of the fluid dynamics. Under what circumstances can
the equations of motion for a solution of colloids
suspended in a solvent of much smaller particles be approximated by
the stochastic equations of motion (\ref{eq:langevin2})? Or, equivalently, in
which situations can the time evolution of the probability density function for
the full system of colloid and solvent particles be approximated by the
Smoluchowski equation (\ref{eq:Smol})? In Refs.\
\cite{Resibois,Resibois_DeLeener} these questions are addressed for the case of
a fluid in which one
of the particles is much larger than the rest and when the solvent friction
coefficient $\Gamma^{-1}$ is sufficiently large, that one can argue that
the time evolution of the
probability distribution function for the single big (colloid) particle is
governed by the single particle Smoluchowski equation. Here we present an
argument which is a simplified generalisation
(for $N$ big colloid particles) of this argument. The derivation does not
contain new ideas. However, we do make connections between older, well known,
results concerning the dynamics of colloidal fluids and more recent
developments in the theory of solvent mediated effective potentials between
colloids in solution \cite{Likos}. The following therefore provides much insight
to the physics incorporated in the Smoluchowski equation and therefore also in
the DDFT, and applies generally to colloidal fluids.

We consider a fluid consisting of a single species of $N$ colloid particles of
mass $m$, suspended in a single component
solvent composed of $S$ solvent particles, of mass $M$.
We denote the coordinates of the $i^{\mathrm th}$ colloid by $\rr_i$ and
momentum $\pp_i$. The set of colloid position coordinates we denote by
$\rr^N \equiv \{\rr_1, ...,\rr_N\}$ and similarly
$\pp^N \equiv \{\pp_1, ...,\pp_N\}$. For the solvent particles we denote
the location in phase space of the $\nu^{\mathrm th}$ solvent particle
by $(\RR_{\nu},\PP_{\nu})$. Just as for the colloids, we denote the sets
$\RR^S \equiv \{\RR_1, ...,\RR_S\}$ and $\PP^S \equiv \{\PP_1, ...,\PP_S\}$.
The Hamiltonian for this system is:
\begin{eqnarray}
{\cal H}(\rr^N,\pp^N,\RR^S,\PP^S,t) &= \frac{1}{2m} \sum_{i=1}^N | \pp_i |^2
+\frac{1}{2M} \sum_{\nu=1}^S | \PP_{\nu} |^2 \nonumber \\
&+V_N(\rr^N,t)+V_{N,S}(\rr^N,\RR^S)+V_S(\RR^S,t),
\end{eqnarray}
where the first two terms on the right hand side are the colloid and solvent
kinetic energy contributions to the Hamiltonian. $V_N(\rr^N,t)$ is the colloid
potential energy:
\begin{eqnarray}
\fl
V_N(\rr^N,t)  = \sum_{i=1}^N v_{ext}(\rr_i,t) 
+ \frac{1}{2} \sum_{j \neq i} \sum_{i=1}^N v_2(\rr_i,\rr_j) 
+ \frac{1}{6} \sum_{k \neq j \neq i}\sum_{j \neq i} \sum_{i=1}^N
v_3(\rr_i,\rr_j,\rr_k)  + ...
\label{eq:V_fn}
\end{eqnarray}
which is assumed to be
made up of a one-body term (external potential $v_{ext}(\rr_i,t)$ on
each colloid particle) a two-body term ($v_2(\rr_i,\rr_j)$ is the pair
potential), a three body term $v_3(\rr_i,\rr_j,\rr_k)$, etc.
Note that $V_N(\rr^N,t)$ describes the {\em direct} or bare colloid-colloid
interaction potential; it does not
involve effective solvent mediated interactions and is therefore not the
same as $U_N(\rr^N,t)$ in Eq.\ (\ref{eq:U_fn}) -- we will further
clarify this issue later (see Eq.\ (\ref{eq:new_U_fn}))
\footnote{For generality we assume $V_N(\rr^N,t)$ contains higher body terms,
although it is generally assumed that the bare
interactions between particles are pairwise with no higher body terms.
In principle, the higher-body terms arise from integrating over quantal degrees
of freedom in obtaining $V_N(\rr^N,t)$.}.
Similarly, $V_S(\RR^S,t)$ is the solvent potential energy and
$V_{N,S}(\rr^N,\RR^S)$ is the potential energy arising from interactions between
the colloid and the solvent particles.

We can define a phase space probability density function
$f^{(N+S)}(\rr^N,\pp^N,\RR^S,\PP^S,t)$ for the fluid, and its  time evolution is
governed by the (exact) Liouville equation \cite{HM}:
\begin{eqnarray}
\fl
\frac{\partial f^{(N+S)}}{\partial t}
+\frac{1}{m} \sum_{i=1}^N \pp_i \cdot \frac{\partial f^{(N+S)}}{\partial \rr_i}
+\sum_{i=1}^N \XX_i \cdot \frac{\partial f^{(N+S)}}{\partial \pp_i}
+\frac{1}{M} \sum_{\nu=1}^S \PP_{\nu} \cdot
\frac{\partial f^{(N+S)}}{\partial \RR_{\nu}}
 \nonumber \\
+\sum_{\nu=1}^S \YY_{\nu} \cdot \frac{\partial f^{(N+S)}}{\partial \PP_{\nu}}
+\sum_{\nu=1}^S \bar{\ZZ}_{\nu} \cdot
\frac{\partial f^{(N+S)}}{\partial \PP_{\nu}}
+\sum_{i=1}^N \ZZ_i \cdot \frac{\partial f^{(N+S)}}{\partial \pp_i}=0,
\label{eq:Liouville}
\end{eqnarray}
where
\begin{eqnarray}
 \XX_i=-\frac{\partial V_N}{\partial \rr_i} \, , \hspace{.5cm}
 \YY_{\nu}=-\frac{\partial V_S}{\partial \RR_{\nu}} \, , \hspace{.5cm}
\ZZ_i=-\frac{\partial V_{N,S}}{\partial \rr_i} \, , \hspace{.5cm}
{\bar\ZZ}_{\nu}=-\frac{\partial V_{N,S}}{\partial \RR_{\nu}} \, ,
\label{eq:forces}
\end{eqnarray}
are forces on the particles. The Liouville equation is a statement of the
continuity of $f^{(N+S)}$ in phase space over time.

Since the solvent particles are much smaller than the colloids, they will
equilibrate on a time scale $\tau_s$ much smaller than the time scale $\tau_c$
on which the colloids equilibrate, i.e.\ $\tau_c \gg \tau_s$. We are
interested in phenomena that occur on
time scales $\sim \tau_c$, so we can assume that effectively the
solvent particles equilibrate instantaneously, and when we integrate over the
solvent degrees of freedom in Eq.\ (\ref{eq:Liouville})
we obtain an equation similar to the Liouville
equation for a one component fluid of $N$ particles, but with an additional
`solvent' term due to interactions between the colloid and solvent particles:
\begin{eqnarray}
\frac{\partial f^{(N)}}{\partial t}
+\frac{1}{m} \sum_{i=1}^N \pp_i \cdot \frac{\partial f^{(N)}}{\partial \rr_i}
+\sum_{i=1}^N \XX_i \cdot \frac{\partial f^{(N)}}{\partial \pp_i}
= \left( \frac{\partial f^{(N)}}{\partial t} \right)_{solvent},
\label{eq:colloid_Liouville}
\end{eqnarray}
where the colloid reduced probability density function is
\begin{eqnarray}
f^{(N)}(\rr^N,\pp^N,t) \equiv \int \dP^S \int \dR^S
f^{(N+S)}(\rr^N,\pp^N,\RR^S,\PP^S,t)
\end{eqnarray}
and formally the `solvent' term is
\begin{eqnarray}
\left( \frac{\partial f^{(N)}}{\partial t} \right)_{solvent}
\equiv -\sum_{i=1}^N\frac{\partial}{\partial \pp_i} \cdot \int \dP^S \int \dR^S
\ZZ_i f^{(N+S)}.
\label{eq:colloid_solvent_term}
\end{eqnarray}
The term $\int \dP^S \int \dR^S \ZZ_i f^{(N+S)}$ in
(\ref{eq:colloid_solvent_term}) is proportional to
the average force exerted on colloid $i$ by
the solvent particles. On this `coarse-grained' time scale $\gg \tau_s$,
the leading order contributions to this force are
a one body (Stokes) drag force on each colloid, $-\gamma \pp_i$, where $\gamma$
is a friction coefficient, and a force term,
$\xx_i$, due to solvent mediated interactions between the colloids. This force
is
\begin{eqnarray}
\xx_i=-\frac{\partial \Phi_N}{\partial \rr_i},
\label{eq:SM_forces}
\end{eqnarray}
where $\Phi_N(\rr^N)$ is the effective solvent mediated potential between the
colloid particles. For an
equilibrium fluid $\Phi_N(\rr^N)=-k_BT \ln Q_S(\rr^N)$, where $Q_S(\rr^N)$ is a
partial partition function \cite{Likos}:
\begin{eqnarray}
Q_S(\rr^N)= \frac{\Lambda_S^{-3S}}{S!} \int \dR^S \exp \left[ - \frac{1}{k_BT}
\left( V_{N,S}(\rr^N,\RR^S)+V_S(\RR^S,t) \right) \right],
\end{eqnarray}
and $\Lambda_S$ is the de-Broglie wavelength for the solvent particles.
For the equilibrium fluid $\Phi_N(\rr^N)$ can, in principle, be calculated. 
In general
\begin{eqnarray}
\fl
\Phi_N(\rr^N)=\tilde{V}\phi_0+\sum_{i=1}^N \phi_1(\rr_i)+
\frac{1}{2} \sum_{j \neq i} \sum_{i=1}^N \phi_2(\rr_i,\rr_j) 
+ \frac{1}{6} \sum_{k \neq j \neq i}\sum_{j \neq i} \sum_{i=1}^N
\phi_3(\rr_i,\rr_j,\rr_k)  + ...,
\label{eq:SM_pot}
\end{eqnarray}
where $\tilde{V}$ is the volume of the system, $\phi_0$ is a zero-body potential
\cite{Likos}, $\phi_1(\rr_i)$ is the solvent mediated one-body potential (for
example with the fluid container walls), $\phi_2(\rr_i,\rr_j)$ is a two body
pair potential,
$\phi_3(\rr_i,\rr_j,\rr_k)$ is a three body potential, and so on. These
potentials are generally density dependent.
We shall assume that the non-equilibrium solvent mediated potential
is the same as the equilibrium solvent mediated potential (\ref{eq:SM_pot}).
This approximation should be reliable on time scales $\gg \tau_s$, since on
these time scales the solvent particles are effectively at equilibrium.
Thus the leading order terms in a Taylor expansion of
the force term in Eq.\ (\ref{eq:colloid_solvent_term}) can be expressed as:
\begin{eqnarray}
\left( \frac{\partial f^{(N)}}{\partial t} \right)_{solvent}
\simeq \sum_{i=1}^N \frac{\partial}{\partial \pp_i} \cdot
\left( (\gamma \pp_i-\xx_i) f^{(N)}
+\theta \frac{\partial f^{(N)}}{\partial \pp_i}+...\right),
\label{eq:colloid_solvent_term_Taylor}
\end{eqnarray}
where $\theta$ is a mobility coefficient, which, in principle, is a function of
the colloidal phase space coordinates. However, we shall assume $\gamma$ and
$\theta$ to be
constants. If we now substitute Eq.\ (\ref{eq:colloid_solvent_term_Taylor}) into
Eq.\ (\ref{eq:colloid_Liouville}), retaining only the two leading order terms,
then we obtain:
\begin{eqnarray}
\frac{\partial f^{(N)}}{\partial t}
+\frac{1}{m} \sum_{i=1}^N \pp_i \cdot \frac{\partial f^{(N)}}{\partial \rr_i}
+\sum_{i=1}^N (\XX_i+\xx_i) \cdot \frac{\partial f^{(N)}}{\partial \pp_i}
\nonumber \\
\hspace{1cm}
= \gamma \sum_{i=1}^N \frac{\partial}{\partial \pp_i} \cdot \pp_i f^{(N)}
+ \theta \sum_{i=1}^N \frac{\partial^2 f^{(N)}}{\partial \pp_i^2}.
\label{eq:kramers}
\end{eqnarray}
We recognise this as the Kramers equation -- i.e.\ a generalised Fokker-Planck
equation \cite{risken84}. At this point we can also make the
connection between the constants
$\gamma$ and $\theta$ via the fluctuation dissipation theorem, and we find that
$\theta=m k_BT \gamma$. In other words, when $\theta=m k_BT \gamma$,
the {\em equilibrium} colloid reduced probability distribution function is
correctly given by
\begin{eqnarray}
f_0^{(N)} \propto \exp \left[ - \beta \left( \frac{1}{2m} \sum_{i=1}^N
| \pp_i |^2 +V_N(\rr^N)+\Phi_N(\rr^N) \right) \right],
\label{eq:equilib_fN}
\end{eqnarray}
where $V_N(\rr^N)$, the term describing the direct interactions between the
colloids, is given by Eq.\ (\ref{eq:V_fn}) with a time independent external
potential.

We now clarify the difference between the potentials $U_N(\rr^N,t)$ in Eq.\
(\ref{eq:U_fn}) and $V_N(\rr^N,t)$ in Eq.\ (\ref{eq:V_fn}).
We see that the total colloid potential $U_N(\rr^N,t)$,
is given by (see Eqs.\ (\ref{eq:V_fn}) and (\ref{eq:SM_pot})):
\begin{eqnarray}
U_N(\rr^N,t)=V_N(\rr^N,t)+\Phi_N(\rr^N).
\label{eq:new_U_fn}
\end{eqnarray}
i.e., we see explicitly here that $U_N(\rr^N,t)$
is the sum of the direct colloid interaction potential (\ref{eq:V_fn})
and the solvent mediated potential (\ref{eq:SM_pot}).
In Eq.\ (\ref{eq:U_fn}) we can therefore identify the
total colloid effective one body potential
$u_{ext}(\rr_i)=v_{ext}(\rr_i)+\phi_1(\rr_i)$, as the sum of a direct
potential and a solvent mediated potential. Similarly, the total colloid
effective pair potential
$u_2(\rr_i,\rr_j)=v_2(\rr_i,\rr_j)+\phi_2(\rr_i,\rr_j)$, the total colloid
effective three body potential
$u_3(\rr_i,\rr_j,\rr_k)=v_3(\rr_i,\rr_j,\rr_k)+\phi_3(\rr_i,\rr_j,\rr_k)$,
and the higher body effective potentials are a sum of a direct contribution and
a solvent mediated contribution.

The Kramers equation (\ref{eq:kramers}) can also be obtained as the
(generalised) Fokker-Planck equation for the time evolution of the probability
density function for $N$ colloid particles with
the following {\em stochastic} equations
of motion \cite{risken84}:
\begin{eqnarray}
\frac{\dr_i}{{\mathrm d} t}=\frac{\pp_i}{m},\nonumber \\
\frac{{\mathrm d} \pp_i}{{\mathrm d} t}=-\gamma \pp_i
-\frac{\partial}{\partial \rr_i}(V_N(\rr^N,t)+\Phi_N(\rr^N))+\GG_i(t),
\label{eq:langevin1}
\end{eqnarray}
where $\GG_i(t)=(\xi_i^x(t),\xi_i^y(t),\xi_i^z(t))$ is a white noise term with
correlations given by Eq.\ (\ref{eq:noise_term}).
In a stochastic treatment of the colloidal fluid one considers the above
equations of motion, (\ref{eq:langevin1}), solely for the
colloids and one incorporates the effect of the solvent via the stochastic noise
term, the friction coefficient $\gamma$ and the solvent mediated potential
$\Phi_N(\rr^N)$. In the treatment above, we arrived at the Kramers
equation (\ref{eq:kramers}) by approximating the Liouville
equation for the full system of solvent and colloid particles.

In deriving the Kramers equation (\ref{eq:kramers}) it is clear that we made a
very large
reduction in number of degrees of freedom in the description of the fluid,
since we have integrated over the solvent degrees of freedom. However, there are
still $6N$ degrees of freedom in (\ref{eq:kramers}) and further reductions are
required. We now focus on the colloid probability density function
\begin{eqnarray}
P^{(N)}(\rr^N,t) = \int \dpp^N f^{(N)}(\rr^N,\pp^N,t).
\end{eqnarray}
In order to obtain an equation for the time evolution of $P^{(N)}(\rr^N,t)$ we
have to make further approximations.
The following is a generalisation of the ``quick and dirty'' approach in Ref.\
\cite{BH} (see also Refs.\ \cite{risken84,BocquetAmJP1997}).
On integrating with respect to the colloid momentum degrees of freedom in
Eq.\ (\ref{eq:kramers}), we obtain the following continuity equation:
\begin{eqnarray}
\frac{\partial P^{(N)}}{\partial t}
+\sum_{i=1}^N \frac{\partial}{\partial \rr_i} \cdot \JJ_i = 0,
\label{eq:continuity}
\end{eqnarray}
where the current
\begin{eqnarray}
\JJ_i \equiv \int \dpp^N \frac{1}{m} \pp_i f^{(N)}(\rr^N,\pp^N,t),
\label{eq:J}
\end{eqnarray}
and we have used the fact that the integrals $\int \dpp_i (\partial f^{(N)}/
\partial \pp_i)$, $\int \dpp_i (\partial^2 f^{(N)}/ \partial \pp_i^2)$ and
$\int \dpp_i (\partial (\pp_if^{(N)})/ \partial \pp_i)$ are all equal to zero.

Also, if we multiply Eq.\ (\ref{eq:kramers}) through by $\pp_k/m$, the velocity
of the $k^{{\mathrm th}}$ colloid, and then
integrate over all the colloid momentum degrees of freedom, we obtain:
\begin{eqnarray}
\fl
\frac{\partial \JJ_k}{\partial t}
+\sum_{i=1}^N \int \dpp^N \frac{\pp_k}{m}
\left( \frac{\pp_i}{m} \cdot \frac{\partial f^{(N)}}{\partial \rr_i} \right)
+\sum_{i=1}^N  \int \dpp^N \frac{\pp_k}{m}
\left( (\XX_i+\xx_i) \cdot \frac{\partial f^{(N)}}{\partial \pp_i} \right)
\nonumber \\
= \gamma \sum_{i=1}^N \int \dpp^N \frac{\pp_k}{m}
\frac{\partial}{\partial \pp_i} \cdot \pp_i f^{(N)}
- \gamma mk_BT \sum_{i=1}^N \int \dpp^N \frac{\pp_k}{m}
\frac{\partial^2 f^{(N)}}{\partial \pp_i^2}.
\label{eq:dJ_by_dt}
\end{eqnarray}
If we now make a `local momentum equilibrium' approximation \cite{BH}, which
sets terms such as
$\int \dpp^N p_{i,a}p_{j,b} f^{(N)}= mk_BT P^{(N)}\delta_{i,j}
\delta_{a,b}$, where $p_{i,a}$ is the $a$--component ($a,b=x,y,z$)
of the momentum of the $i^{{\mathrm th}}$ particle,
then we find that Eq.\ (\ref{eq:dJ_by_dt}) reduces to
\begin{eqnarray}
\frac{\partial \JJ_k}{\partial t}
+\frac{k_BT}{m} \frac{ \partial P^{(N)}}{\partial \rr_k}
- \frac{1}{m} (\XX_k+\xx_k) P^{(N)}= - \gamma \JJ_k.
\label{eq:dJ_by_dt_b}
\end{eqnarray}
If the friction constant $\gamma$ is sufficiently large then we can neglect the
first term in (\ref{eq:dJ_by_dt_b}), as it will be negligible compared to the
friction term $- \gamma \JJ_k$, on the Brownian time scales $\tau_B$,
and we obtain:
\begin{eqnarray}
\JJ_k\simeq -\Gamma k_BT \frac{ \partial P^{(N)}}{\partial \rr_k}
- \Gamma \frac{\partial U_N}{\partial \rr_k} P^{(N)},
\label{eq:J_2}
\end{eqnarray}
where $\Gamma=1/m\gamma$. Substituting (\ref{eq:J_2}) into Eq.\
(\ref{eq:continuity}), we obtain the Smoluchowski equation (\ref{eq:Smol}).

The Smoluchowski equation is generally presented from the stochastic viewpoint,
starting from the Langevin equations of motion (\ref{eq:langevin2}).
However, as shown above, one can argue for its use as an approximation to the
exact Liouville equations for the time evolution of the probability density
function for colloidal fluids. The above derivation demonstrates the physical
conditions under which the time evolution of the probability density function of
a fluid of colloidal particles is described by the Smoluchowski equation. By
going to the Smoluchowski equation (with an effective potentials between the
colloids), the
description of the fluid is reduced to one based on only the position
coordinates of the colloids, rather than the full set of phase space
coordinates for the colloid and solvent particles, which is, of course, a
significant simplification.

\section{Discussion and conclusions}
\label{sec:discussion}

In Sec.\ \ref{sec:theDDFT} we introduced the DDFT of Marconi and
Tarazona \cite{Marconi:TarazonaJCP1999,Marconi:TarazonaJPCM2000} for colloidal
fluids, following a recent derivation in Ref.\ \cite{Archer7} of the DDFT
starting from the Smoluchowski equation.
We emphasise that this is an approximate theory.
In Sec.\ \ref{sec:application} we applied the DDFT to the specific case of a
binary fluid of GCM particles confined in a spherical cavity of variable size.
We find that the DDFT is able to describe accurately the time evolution of the
highly structured fluid one-body density profiles. We find a variety of
collective dynamic processes that are accounted for by the DDFT. For example,
the case in Fig.\ \ref{fig:5}, where we see a particle concentration `wave'
travelling through the fluid into the centre of the cavity
that allows the fluid to
equilibrate into the phase-separated configuration slightly
faster than one would
expect, bearing in mind the results displayed in Fig.\ \ref{fig:4}, which showed
that the fluid-fluid interface persists for a long time after the cavity radius
${\cal R}$ is increased. Such phenomena are very difficult to simulate because
of the large number of particles
involved. It is in situations such as these where the DDFT should provide a
useful tool to analyse the fluid dynamics.

Our strategy is to first consider cases of only a few hundred particles where
BD simulations can be used to assess the accuracy of the DDFT results.
In these test cases with fewer particles, where the fluid exhibits a limited
degree of phase separation and wetting behaviour,
we find excellent agreement between
the DDFT and the BD simulation results (see for example Figs.\
\ref{fig:1} and \ref{fig:2}). This gives us confidence to trust the reliability
of the DDFT when applied to cases with many more particles, where comparison
with simulation becomes computationally too expensive.
We predict that were one to perform a BD simulation for (say)
the cases presented in Figs.\ \ref{fig:4} and
\ref{fig:5}, one would find just as good agreement between the DDFT results and
BD simulation results. In fact, one should argue that since the total
fluid densities are higher in these situations, and the fact that the RPA
functional
(\ref{eq:F_ex}) becomes more accurate at higher densities \cite{Likos,Archer1},
the results from the DDFT should be even more reliable in these cases.

In Sec.\ \ref{sec:LKS}, we derive the Smoluchowski equation,
starting from the exact
Liouville equations by making a series of approximations in order to go from a
description of the fluid in terms of the full set of colloid and solvent phase
space coordinates to a (far more manageable) description solely in terms of the
colloid position coordinates (the Smoluchowski equation (\ref{eq:Smol})). Such
integration over
degrees of freedom is a process that is required in order to render a practical
statistical description of any condensed matter system. We believe our
derivation is useful because it highlights some of the approximations involved
in the Smoluchowski equation for colloids and therefore also in the DDFT and
thus sheds light on the status of the DDFT.

The Smoluchowski equation is usually presented from the stochastic viewpoint,
since it is the Fokker-Planck equation for the Langevin equations
(\ref{eq:langevin2}). This connection makes it clear that the Smoluchowski
equation and thus the DDFT neglect hydrodynamic effects. To include
hydrodynamic effects one would have to treat the solvent at a level beyond that
taken here, where the effects of the solvent are incorporated in the DDFT via
the effective solvent mediated potential between the colloids and in the
friction coefficient $\Gamma^{-1}$. One can build in hydrodynamic effects in
the Smoluchowski equation by replacing
$\Gamma$ with a matrix $\Gamma_{ij}$, describing the hydrodynamic
coupling between colloids $i$ and $j$
\cite{Archer7,PuseyTough,VerbergetalPRE2000}. However, this makes the
reduction of the Smoluchowski equation to a DDFT much less straightforward.

We conclude that the DDFT, Eq.\ (\ref{eq:mainres_multi}), should provide a good
theory for the dynamics of the one body density of a colloidal fluid, even when
there is phase separation and wetting of interfaces, provided
that i) the friction constant characterising the solvent, $\Gamma^{-1}$, is
large enough and ii) there exists an accurate approximation for the
excess Helmholtz free energy functional ${\cal F}_{ex}[\rho]$.
Many colloidal fluids can be accurately modelled as effective hard-sphere
fluids and therefore we expect that the DDFT combined with 
fundamental measure free energy functionals \cite{RosenfeldPRL1989,white_bear}
for hard spheres to provide a reliable theory for the dynamics of the one body
density profile of a (hard-sphere) colloidal fluid.

\ack
I am grateful to Christos Likos and Bob Evans for useful discussions and
a critical reading the manuscript. I am also
grateful to Christos Likos for providing a static BD simulation code.
I acknowledge the support of EPSRC under grant number GR/S28631/01.


\section*{References}
\begin{thebibliography}{99}

\bibitem{opticaltweezers}
See for example Molloy J E and Padgett M J 2002 {\it Contemporary Phys.}
{\bf 43} 241 and references therein

\bibitem{BH}
Barrat J-L and Hansen J-P 2003 {\it Basic Concepts for Simple Liquids},
Cambridge University Press, Cambridge

\bibitem{Likos}
Likos C N 2001 {\it Phys. Rep.} {\bf 348} 267

\bibitem{Roland}
Roth R, Evans R and Dietrich S 2000 {\it Phys. Rev. E} {\bf 62} 5360

\bibitem{risken84}
Risken H 1984 {\it The Fokker-Planck Equation}, Springer, Berlin

\bibitem{Marconi:TarazonaJCP1999}
Marini Bettolo Marconi U and Tarazona P 1999 {\it J. Chem. Phys.} {\bf 110} 8032

\bibitem{Marconi:TarazonaJPCM2000}
Marini Bettolo Marconi U and Tarazona P 2000 {\it J. Phys.: Condens. Matter}
{\bf 12} A413

\bibitem{Evans79}
Evans R 1979 {\it Adv. Phys.} {\bf 28} 143

\bibitem{Archer7}
Archer A J and Evans R 2004 {\it J. Chem. Phys.} {\bf 121} 4246

\bibitem{Archer8}
Archer A J and Rauscher M 2004 {\it J.Phys. A: Math. Gen.} {\bf 37} 9325

\bibitem{joe:christos}
Dzubiella J and Likos C N 2003 {\it J. Phys.: Condens. Matter} {\bf 15} L147

\bibitem{Dautenhahn}
Dautenhahn J and Hall C K 1994 {\it Macromolecules} {\bf 27} 5399

\bibitem{LouisetalPRL2000}
Louis A A, Bolhuis P G, Hansen J-P and Meijer E J 2000 {\it Phys. Rev. Lett.}
{\bf 85} 2522

\bibitem{BolhuisetalJCP2001}
Bolhuis P G, Louis A A, Hansen J-P, and Meijer E J 2001 {\it J. Chem. Phys.}
{\bf 114} 4296

\bibitem{LouisetalPhysicaA2002}
Louis A A, Bolhuis P G, Finken R, Krakoviack V, Meijer E J and Hansen J-P 2002
{\it Physica A} {\bf 306} 251

\bibitem{BolhuisLouisMacrom2002}
Bolhuis P G and Louis A A 2002 {\it Macromolecules} {\bf 35} 1860

\bibitem{cnl:jcp}
Likos C N, Rosenfeldt S, Dingenouts N, Ballauff M, Lindner P, Werner N,
and V\"ogtle F 2002 {\it J. Chem. Phys.} {\bf 117} 1869

\bibitem{cnl:macrom}
Likos C N, Schmidt M, L\"owen H, Ballauff M, P\"otschke D, and Lindner P 2001
{\it Macromolecules} {\bf 34} 2914

\bibitem{LangJPCM}
Lang A, Likos C N, Watzlawek M and L\"owen H, {\it J. Phys.: Condens. Matter}
{\bf 12} 5087

\bibitem{paper1}
Louis A A, Bolhuis P G and Hansen J-P 2000 {\it Phys. Rev. E} {\bf 62} 7961

\bibitem{Archer1}
Archer A J and Evans R 2001 {\it Phys. Rev. E} {\bf 64} 041501

\bibitem{Archer2}
Archer A J and Evans R 2002 {\it J. Phys.: Cond. Matter} {\bf 14} 1131

\bibitem{Archer6}
Archer A J, Likos C N and Evans R 2004 {\it J.Phys.: Cond. Matter} {\bf 16} L297

\bibitem{Patrykiejewetal2004}
Patrykiejew A, Pizio O and Sokolowski S 2004 {\it Mol. Phys.} {\bf 102} 801

\bibitem{flor2}
Penna F and Tarazona P 2003 {\it J. Chem. Phys.} {\bf 68} 1766

\bibitem{flor1}
Penna F, Dzubiella J and Tarazona P 2003 {\it Phys. Rev. E} {\bf 68} 061407

\bibitem{RexetalPRE2005}
Rex M, L\"owen H, and Likos C N 2004 submitted

\bibitem{Evans92}
Evans R 1992, in {\it Fundamentals of Inhomogeneous Fluids}, ed.
Henderson D, Dekker, New York, ch.\ 3

\bibitem{HM}
Hansen J-P and McDonald I R 1986 {\it Theory of Simple Liquids}, Academic,
London, 2nd ed.

\bibitem{Dhont_book}
Dhont J K G 1996 {\it An Introduction to Dynamics of Colloids}, Elsevier,
Amsterdam

\bibitem{DhontJCP1996}
Dhont J K G 1996 {\it J. Chem. Phys.} {\bf 105} 5112

\bibitem{RowlinsonWidom}
Rowlinson J S and Widom B 1982 {\it Molecular Theory of Capillarity}, Oxford
University Press, Oxford

\bibitem{Archer3}
Archer A J, Evans R and Roth R 2002 {\it Europhys. Lett.} {\bf 59} 526

\bibitem{Archer5}
Archer A J and Evans R 2003 {\it J. Chem. Phys.} {\bf 118} 9726

\bibitem{Archer10}
Archer A J, Evans R, Roth R and Oettel M 2005 {\it J. Chem. Phys} accepted,
{\it cond-mat/0411557}

\bibitem{AllenTildesley}
Allen M P and Tildesley T J 1989 {\it Computer Simulation of Liquids}, Oxford:
Clarendon

\bibitem{Resibois}
R\'esibois P 1968 {\it Electrolyte Theory}, Harper and Row, New York

\bibitem{Resibois_DeLeener}
R\'esibois P and DeLeener M 1977 {\it Classical Kinetic Theory of Fluids},
John Wiley, New York

\bibitem{BocquetAmJP1997}
Bocquet L 1997 {\it Am. J. Phys.} {\bf 65} 140

\bibitem{PuseyTough}
Pusey P N and Tough R J A 1985, in {\it Dynamic Light Scattering}, ed. Pecora R,
Plenum Press, New York

\bibitem{VerbergetalPRE2000}
Verberg R, de Schepper I M and Cohen E G D 2000 {\it Phys. Rev. E} {\bf 61} 2967

\bibitem{RosenfeldPRL1989}
Rosenfeld Y 1989 {\it Phys. Rev. Lett.} {\bf 63} 980

\bibitem{white_bear}
Roth R, Evans R, Lang A and Kahl G 2002 {\it J. Phys. Cond. Matter} {\bf 14}
12063

\end{thebibliography}
\end{document}